\newcommand\ignore[1]{}

\documentclass[11pt, oneside]{article}   	
\usepackage{geometry}                		
\usepackage[parfill]{parskip}    		
\usepackage{graphicx}				
\usepackage{color, colortbl}								
\usepackage[left]{lineno}
\usepackage{amssymb}
\usepackage{amsmath}
\usepackage{amssymb}
\usepackage{graphicx}
\usepackage{hyperref}
\usepackage{float}
\usepackage{fullpage}
\usepackage{braket}
\usepackage{url}
\usepackage{cite}
\usepackage{subcaption}
\usepackage{upgreek}
\usepackage{outlines}
\usepackage{braket}
\usepackage{appendix}
\usepackage{xspace}
\usepackage{tikz}
\usepackage{tikz-cd}
\usepackage{placeins}
\usepackage{multirow}
\usepackage{sidecap}
\usepackage{soul}
\usepackage{enumitem}

\usetikzlibrary{matrix,arrows,decorations.pathmorphing}
\usetikzlibrary{patterns}
\usetikzlibrary{trees}
\usetikzlibrary{decorations.pathmorphing}
\usetikzlibrary{decorations.markings}
\usetikzlibrary{positioning,arrows}
\usetikzlibrary{decorations.pathreplacing}
\usetikzlibrary{decorations.pathmorphing}
\usetikzlibrary{decorations.markings}

\usepackage{etoolbox} 

\newcommand*\linenomathpatch[1]{%
	\cspreto{#1}{\linenomath}%
	\cspreto{#1*}{\linenomath}%
	\csappto{end#1}{\endlinenomath}%
	\csappto{end#1*}{\endlinenomath}%
}

\linenomathpatch{equation}
\linenomathpatch{gather}
\linenomathpatch{multline}
\linenomathpatch{align}
\linenomathpatch{alignat}
\linenomathpatch{flalign}

\newcommand\be{\begin{equation}}
\newcommand\ee{\end{equation}}
\newcommand\bea{\begin{eqnarray}}
\newcommand\eea{\end{eqnarray}}

\newcommand{\bdm}{\begin{displaymath}}
\newcommand{\edm}{\end{displaymath}}
\newcommand\nn{ \nonumber\\}

\numberwithin{equation}{section}
\numberwithin{figure}{section}

\renewcommand{\epsilon}{\varepsilon}

\usepackage{authblk}

\renewcommand{\phi}{\varphi}

\graphicspath{{/}{fig/}{rootFigs/}}

\title{Lorentzian OPE Inversion Formula: A Geometric Perspective}

\author[*]{Pulkit Agarwal}
\author[**]{Richard Brower}
\author[+]{Timothy Raben}
\author[++]{Chung-I Tan}
\affil[*]{Centro de Física do Porto, Faculdade de Ci\^{e}ncias da Universidade do Porto, Portugal}
\affil[**]{Boston University, Boston, MA 02215}
\affil[+]{Michigan State University} 
\affil[++]{Brown University, Providence, RI 02912}

\begin{document}
\ignore{\count\footins = 1000
\title{Lorentzian OPE Inversion Formula: A Geometric Perspective}

\author{Pulkit Agarwal}
\affiliation{Centro de Física do Porto, Faculdade de Ci\^{e}ncias da Universidade
	do Porto, Portugal}
\author{Richard C. Brower}
\affiliation{Boston University, Boston, MA 02215}
\author{Timothy G. Raben}
\affiliation{Michigan State University, East Lansing, MI, 48824} 
\author{Chung-I Tan}
\affiliation{Brown University, Providence, RI 02912}
}

\maketitle
\thispagestyle{empty}

\begin{abstract}
	
	We give a new perspective on the Lorentzian OPE inversion formula \cite{caron2017analyticity,simmons2018spacetime}, building on \cite{Agarwal:2023xwl}. We introduce an ``auxiliary'' fourpoint function that can be related to the traditionally defined ones via a Radon transform. The Mellin amplitudes associated with this auxiliary function can be shown to be equivalent to the conventional partial wave amplitudes. This has the intuitive geometrical meaning of a generalization of the Projection-Slice Theorem.
\end{abstract}

\newpage
\thispagestyle{empty}
\tableofcontents
\newpage

\section{Introduction}
\pagenumbering{gobble}
\pagenumbering{arabic}

The fourpoint function plays a central role in CFT studies. Conformal symmetry is highly restrictive and these fourpoint functions are known to reduce to a function of two variables known as the cross ratios (up to kinematic prefactors). When expressed in a conformal block expansion, the partial wave amplitudes associated with these functions can be shown to relate directly to CFT data.~\footnote{The singularity structure of the partial wave amplitudes gives us the spectrum of operators present in the theory, while the residues at these singularities are related to the strength of the conformal three point functions.} Much in analogy with standard Regge theory (or more generally in harmonic analysis), the blocks themselves are understood as eigenfunctions of the conformal Casimir equation. These blocks are completely determined by the kinematics, whereas the amplitudes contain all the dynamical information of the theory \cite{Dolan:2000ut,Dolan:2003hv,dolan2011conformal,Hogervorst:2013sma,qualls2015lectures,Simmons-Duffin:2016gjk,rychkov2017epfl}.

A shift in perspective on this classic story was introduced in \cite{Agarwal:2023xwl}, where instead of thinking about correlation between four points on the spacetime manifold, one considers directly the space on which the \textit{reduced} fourpoint function $f(u,v)$ lives as a manifold in its own right. Going to the cross ratio space from the full conformal group $G$ is, in effect, restricting to a double coset $G//H$ with $H$ an appropriate subgroup~\footnote{For mathematical notations, see \cite{Agarwal:2023xwl}. For the notion of a ``double-coset", i.e., left-right cosets, it is typically denoted by some as $H\backslash G/H$. We prefer using $G//H$, which is also often used in mathematical literature.}. One thus interprets the usual conformal Casimir equation as the Laplace-Beltrami operator \cite{Schomerus2017,Isachenkov2018,Schomerus2018,Buric2020,Buric:2022ucg,agarwal2022application} on this manifold~\footnote{To be more precise, it is the radial component of the Laplace-Beltrami operator, as suggested by the fact that the functions are bi-invariant (see Sec. \ref{sec:groups}).} and its solutions form a basis for the reduced fourpoint function. This basis is made up of {\it zonal spherical functions} on the manifold. An integral representation for these functions appropriate for Lorentzian CFTs was constructed in \cite{Agarwal:2023xwl}.

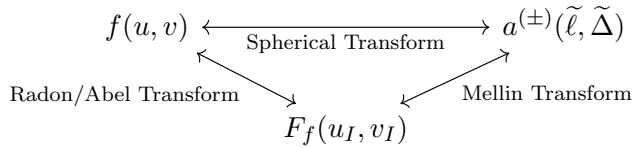
\begin{figure}[h]
	\centering
	\begin{tikzcd}
		{f(u,v)} \arrow[rr, "\text{Spherical Transform}"',leftrightarrow] \arrow[rd, "\text{Radon/Abel Transform}"',leftrightarrow] &    & {a^{(\pm)}(\widetilde\ell,\widetilde\Delta)} \\
		& {F_f(u_I,v_I)} \arrow[ru, "\text{Mellin Transform}"',leftrightarrow] &  
	\end{tikzcd}
	\caption{Tomography of the conformal fourpoint function.}\label{fig:transforms}
\end{figure}

In this paper, we show that this shift in perspective allows further simplification through the notion of  {\it Radon/Abel transform}. (See Fig. \ref{fig:transforms}.)  By introducing a Radon transformed auxiliary fourpoint function, $f(u,v)\rightarrow F_f(u_I,v_I)$, the conventional CFT conformal partial waves can simply be reduced to plane waves and the partial wave amplitudes are a simple Mellin transform, $F_f \rightarrow a^{(\pm)}(\widetilde\ell,\widetilde\Delta)$. In principle, this brings ideas from tomography to CFTs and Fig. \ref{fig:transforms} is the CFT analogue of the well known $FHA$ cycle (relating Fourier, Hankel, and Abel transforms) in Projection-Slice theorems \cite{enwiki:1167004207} which allows us to close the circle for some of the issues discussed in  \cite{Agarwal:2023xwl}.  In particular, this equivalence allows one to address the issue of inversion for causal CFT partial wave amplitudes for non-vanishing {\it double-commutators}, $a^{(\pm)}(\widetilde\ell,\widetilde\Delta)   \leftrightarrow f(u,v)$,  or, equivalently, the so-called {\it double-Discontinuities}~\footnote{An immediate problem is the reconstruction of the full Euclidean fourpoint function from the information obtained from the discontinuity. This is done via dispersion relations \cite{carmi2020conformal} that we shall not discuss in this current work (see the conclusion for some remarks).}. (These are often denoted as ``dDisc" \cite{caron2017analyticity,simmons2018spacetime}. See \cite{Agarwal:2023xwl}  for our precise usage.) 

Conformal blocks, even in the Euclidean setting, are known to not form a complete orthonormal basis. One way this is dealt with is by forming conformal partial waves, which are a linear combination of blocks and shadow blocks \cite{Costa:2012cb}. One of the key distinguishing features between Euclidean and Lorentzian blocks is that, because of the semigroup structure discussed in \cite{Agarwal:2023xwl}, including the shadow is no longer sufficient to construct a complete orthonormal set of partial waves \cite{Raben:2018rbn}. This prevents one from performing the inversion by a standard procedure. Our solution to this problem is, in principle, simple. It can be shown that Mellin transforms associated with the auxiliary functions,  $F_f \rightarrow \widetilde{F}_f$ (see Eq. \ref{eq:mellinF}), are the partial wave amplitudes of the original fourpoint functions $a^{(\pm)}(\widetilde\ell,\widetilde\Delta)$ and inverting Mellin transforms can be carried out more directly. This is our main result.

\begin{figure}[h]
	\centering
	\begin{subfigure}{0.3\textwidth}
		\includegraphics[width=\textwidth]{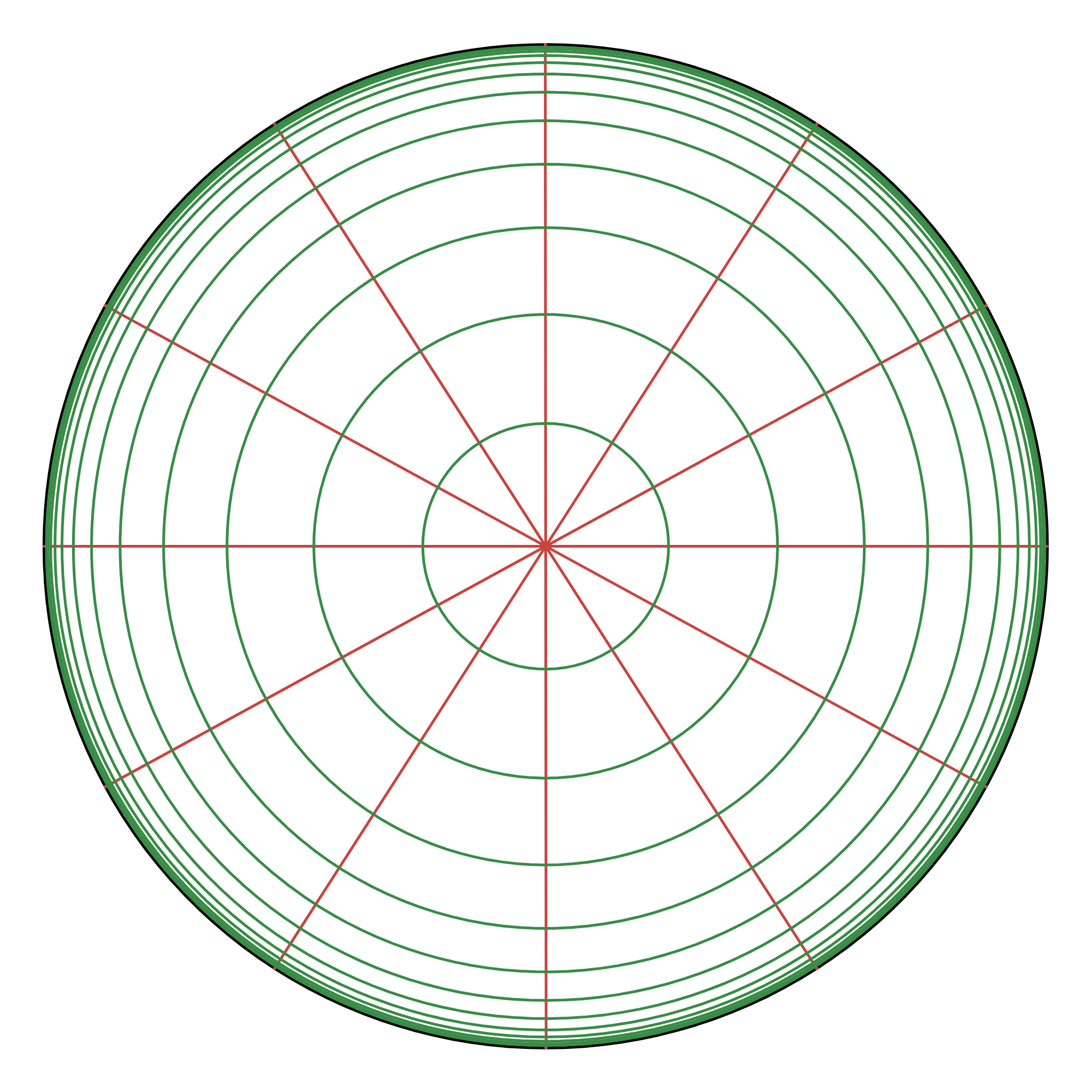}
		\caption{}
	\end{subfigure}
	\quad\quad\quad\quad\quad\quad
	\begin{subfigure}{0.3\textwidth}
		\includegraphics[width=\textwidth]{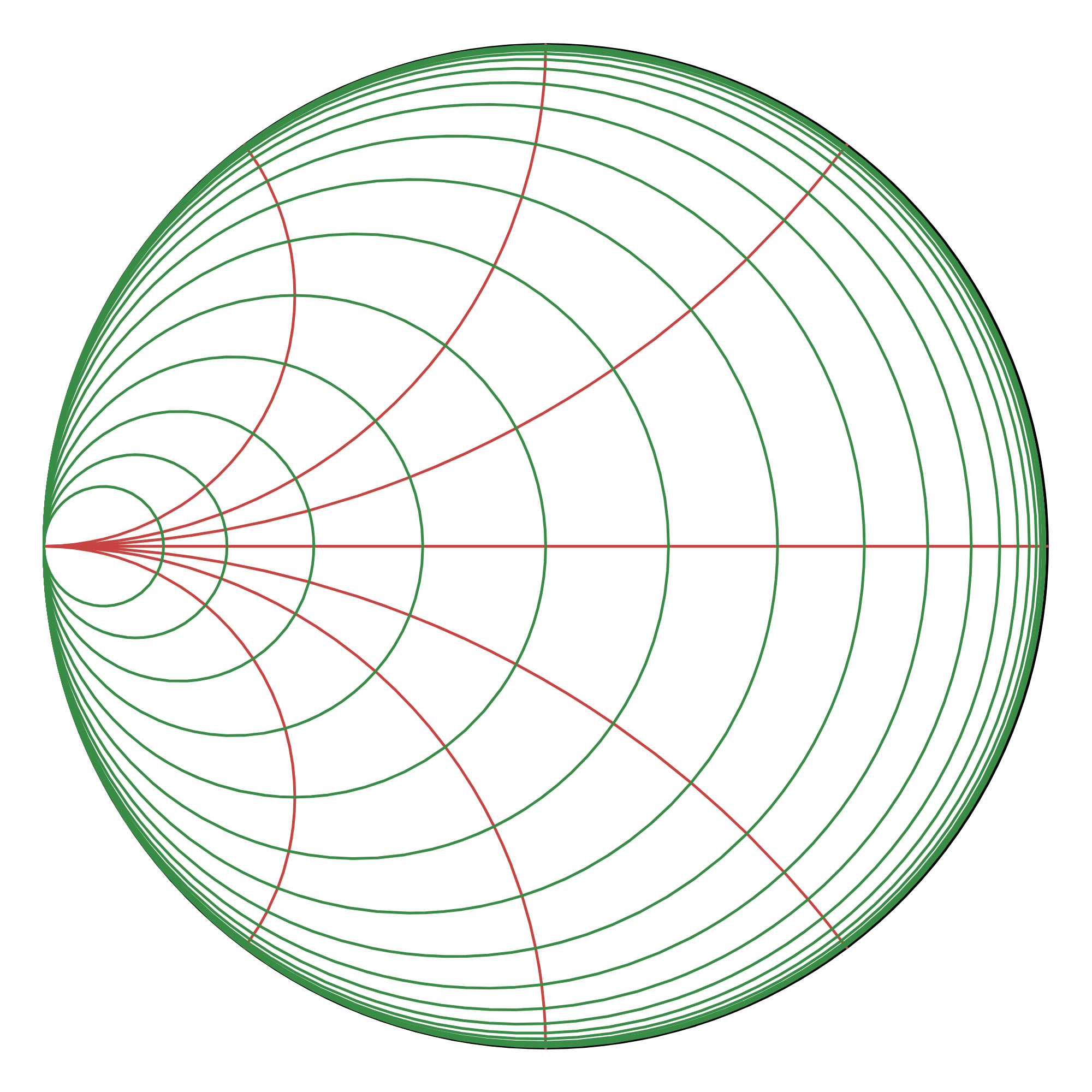}
		\caption{}
	\end{subfigure}
	\caption{Orbits on the Poincar\'e disc model of H$_2$ in two different group parametrizations. Fig. (a) corresponds to the Cartan decomposition where the manifold is identified with coset elements of the form $K(\phi)A(\eta)$. Lines of constant $\eta$ are shown in green whereas lines of constant $\phi$ are shown in red. Fig. (b) is in the Iwasawa decomposition where the coset is parametrized by elements of the form $N(c)A_I(\eta_I)$. Lines of constant $\eta_I$ are in green whereas lines of constant $c$ are in red. All red lines correspond to geodesics. The variable $\eta$ can be shown to parametrize cross ratios on the null boundary. See Sec. \ref{sec:groups} for introduction to group decompositions and App. \ref{app:disc} for explicit coordinates.}
	\label{fig:disc}
\end{figure}

\paragraph{An Illustrated Example:} For CFTs, the construction of the auxiliary functions can be understood intuitively due to conformal bi-invariance \cite{Agarwal:2023xwl}. Consider, for the sake of simplicity, the case of H$_2$, (the upper-half complex plane), which is closely related to CFT$_1$~\footnote{Note that the maximal abelian subgroup of this example has rank 1. For $d$-dimensional Lorentzian CFTs, one typically wants to study the rank 2 case. This is nevertheless an important illustrative example.}. Group theoretically the Poincare disc, Fig. \ref{fig:disc}, is identified with the coset $G/K = SO(2,1)/SO(2)$~\footnote{To denote relevant subgroups of the group $G$ under consideration, we will use the following notations: $K$: maximal compact subgroup; $A$: maximal abelian subgroup; $N$: subgroup of nilpotents.}. We can parametrize the disc in several different ways but there are two specific ways illustrated in Fig. \ref{fig:disc} that will be of particular importance to us. On the left, the disc is parametrized by a radius away from the center and an angle. This is similar to the Euler angle decomposition of a sphere, which is an example of \textit{Cartan decomposition} of the group ($G = KAK$ in group notation). This decomposition is particularly useful for working with $K$-bi-invariance, where functions can be parametrized only by the maximal Abelian subgroup (MASG)  $A$, i.e., when studying ``radial functions". On the right of Fig. \ref{fig:disc}, we parametrize the disc in terms of its horocycles that fix a point on the boundary of the disc. This is an example of \textit{Iwasawa decomposition} of the group ($G = NA_IK$ in group notation) and is convenient for a direct application of Radon/Abel transforms  \cite{lang,sugiura,beerends1987introduction} (or equivalently, Harish-Chandra transforms \cite{lang,gangolli2012harmonic}).

Since our goal is to study CFTs, which reduce to radial functions, we start by considering the class of functions over the disc that are rotationally symmetric, (i.e., bi-$K$ invariance with $K$ compact). Our goal is then to perform harmonic analysis on these symmetric functions. These functions live on the double coset $G//K$ and depend only on the radius away from the center illustrated in Fig. \ref{fig:disc}(a). One can construct a basis of these bi-invariant zonal spherical functions in an integral representation. We start on the coset $G/K$ and integrated out the second copy of $K$ to reduce to the rotationally symmetric space (now parametrized by the maximal Abelian subgroup $A$).

In order to perform inversion, our proposal is effectively a change of variables from Cartan to Iwasawa, and our claim is that it solves the problem of inversion in a transparent way. Instead of integrating over rotations $K$, we integrate out the nilpotents (e.g. translations or special conformal transformations) $N$ first, again arriving at functions over the maximal Abelian subgroup $A_I$. This has a rather nice geometric interpretation. As illustrated in Fig. \ref{fig:disc}, lines of constant $\eta_I$ are parametrized by $n(c)$. They are orthogonal to the geodesics at the point of intersection. The point $(-1,0)$ on the disc is identified with $c\rightarrow \infty$. We are integrating along these transverse lines over $(-\infty,\infty)$. This is precisely the definition of the X-ray transform (or Radon transform in general) as done in standard tomography. The remaining integral over $\eta_I$ required for inversion reduces to a simple Laplace transform (see App. \ref{app:disc} for a full treatment of H$_2$).

Our goal here is to lift this idea into harmonic analysis for CFTs. However, this poses several challenges. First, CFTs live on the null cone in the embedding space. Further, for general $d$, the MASG involved is of rank 2. The mathematical statement for this problem is to perform harmonic analysis on the Grassmannian-like manifold $SO(d,2)/(SO(d-1,1)\times SO(1,1))$. This is a largely open problem, particularly for the ``conformal Lorentzian Grassmannian"~\footnote{The standard Grassmannian would be $SO(d+2)/(SO(d)\times SO(2))$. The Lorentzian Grassmannian would be $SO(d+1,1)/(SO(d)\times SO(1,1))$ describes the null cone of the Lorentz group which is appropriate for Euclidean CFTs. The problem for $SO(2,2)/(SO(2)\times SO(2))$ was dealt with in  \cite{Shilin:2018}.} that we are dealing with here. The goal is to find all required measures for the Radon transform followed by the Laplace transform. After formalizing some general results, we will solve this explicitly for the $d=1,2$ cases.

In Sec. \ref{sec:kinematics}, we give a quick review of the kinematic setup for Lorentzian CFTs. We illustrate the various regions of the cross ratio space and highlight the OPE points of interest. (See Fig.   \ref{fig:uv}.)  In particular, we provide a group theoretically motivated specification of inequivalent holonomies required in defining the desired double-discontinuities.  In Sec. \ref{sec:groups}, we review the group theoretic construction of conformal blocks done in \cite{Agarwal:2023xwl}. In Sec. \ref{sec:inv}, we first introduce Radon transforms and present general results applicable for all spacetime dimensions $d$. We then present explicit calculations for $d=1,2$. We conclude in Sec. \ref{sec:conclusion} with some comments.

\section{Kinematics: From Euclidean to Minkowski Settings}\label{sec:kinematics}
The conformal fourpoint function for scalar fields can be shown to reduce to a scalar function of two cross ratios when it is appropriately normalized, e.g., $
	\braket{\varphi_1(x_1)\varphi_2(x_2)\varphi_3(x_3)\varphi_4(x_4)} = \frac{1}{x^{\Delta_1}_{12}x^{\Delta_3}_{34}} F(u,v)
$,
where we have assumed that the scaling dimension of $\varphi_1$ and $\varphi_2$ is $\Delta_1$ and that of $\varphi_3$ and $\varphi_4$ is $\Delta_3$, with cross ratios defined as $u = x^2_{12}x^2_{34}/x^2_{13}x^2_{24}$ and $v = x^2_{14}x^2_{23}/x^2_{13}x^2_{24}$. The functions $F(u,v)$ admit a conformal block expansion
\begin{equation}
	F(u,v) = \sum_{\ell,\Delta} a(\ell,\Delta) G_{\ell,\Delta}(u,v)
\end{equation}
where $G_{\ell,\Delta}(u,v)$ are called conformal blocks and are solutions of the conformal Casimir equation.  This expansion can be understood in terms of Operator Product Expansions (OPE) where points $x_i$ come close together pairwise, with $G_{\ell,\Delta}(u,v)$ obeying appropriate boundary conditions.

\begin{figure}
	\centering
	\begin{subfigure}{0.4\textwidth}
		\includegraphics[width=\textwidth]{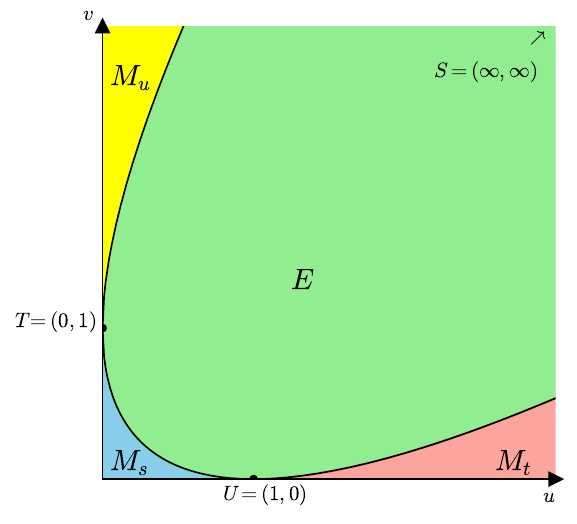}
		\caption{}
	\end{subfigure}
	\quad
		\quad
	\begin{subfigure}{0.4\textwidth}
		\includegraphics[width=\textwidth]{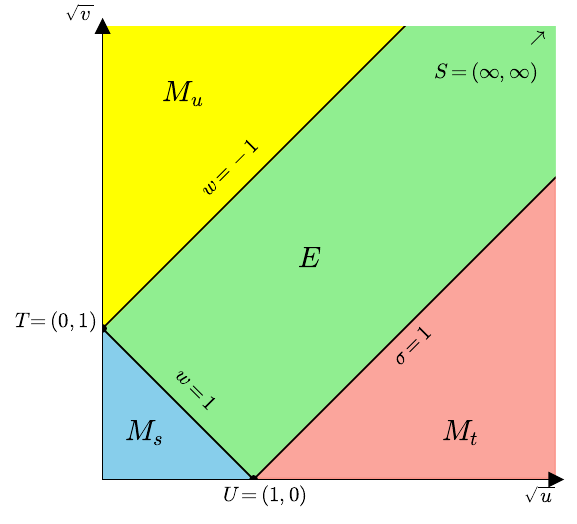}
		\caption{}
	\end{subfigure}
	\quad
	\begin{subfigure}{0.5\textwidth}
		\includegraphics[width=\textwidth]{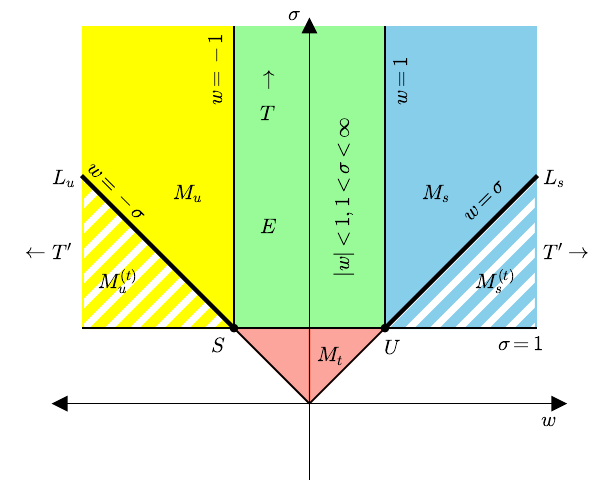}
		\caption{}
	\end{subfigure}
	\caption{The first quadrant of the $ (u,v) $ plane divides into four regions, $E$,  $M_s$, $M_u$ $M_t$, which are multi-sheeted.  In terms of new variables $(w,\sigma)$, Eq. (\ref{eq:wsig}), this is resolved into multple copies. The first quadrant of the $ \sqrt{u},\sqrt{v} $ plane divides into four regions (shown in solid colors) with striaghtline segments.   In terms of new variables $(w,\sigma)$, the corresponding 4 regions are shown in Fig. \ref{fig:uv}(c). Causal Minkowski regions $M^{(t)} _s$, $M^{(t)}_u$,  with non-vanishing double discontinuities are on the second sheet in the $(u,v)$ plane and are now seen in hatched colors.}
	\label{fig:uv}
\end{figure}

In Euclidean CFT, one can consider three different OPE limits leading to three equivalent Euclidean  expansions, $f_E(u,v)$, related by kinematic factors. For simplicity, let us restrict to the case where $\Delta_1=\Delta_3$. Consider the partition $(12)(34)$, leading to the limit where $x_{12}^2,x_{34}^2\rightarrow 0$ or, more accurately, $(u,v)\rightarrow(0,1)$. This is called the $t$-channel OPE. In Fig. \ref{fig:uv}(a), this is labelled by the point $T$. Similarly, one has the $u$-channel OPE limit for $(14)(23)$ where $(u,v)\rightarrow(1,0)$ (labelled by point $U$ in Fig. \ref{fig:uv}(a)) and the $s$-channel OPE limit for $(13)(24)$ where $(u,v)\rightarrow(\infty,\infty)$ (labelled by point $S$ in Fig. \ref{fig:uv}(a)).   Euclidean fourpoint functions live in the green region of Fig. \ref{fig:uv}.  These expansions can be extended {\it analyticially} into Lorentzian settings, (from region $E$ to $M_s$, $M_u$ $M_t$ in Fig. \ref{fig:uv}(a)), while maintaining spacelike separations for all quadratic pairs, i.e., $x_{ij}^2>0$ (we refer to these regions as {\it non-causal}). 

The limit $x_{ij}^2 \rightarrow 0$ can now be approached in three physically distinct ways: they can approach each other similar to the Euclidean case, or they can approach each others' lightcones in a spacelike or timelike separation. This makes the points $S,T,U$ degenerate in the $u$-$v$ plane. To lift this degeneracy, instead of the first quadrant of the $(\sqrt{u},\sqrt{v})$ plane, Fig.  \ref{fig:uv}(b),   it is useful to extend into the whole $(\sqrt{u},\sqrt{v})$, e.g., extending $\sqrt v>0$ to regions where $\sqrt v < 0$. More importantly, the limit involving timelike separation leads to {\it causal OPEs} which are holonomically inequivalent~\footnote{This degeneracy is only somewhat resolved as we shall see shortly (also see Fig. \ref{fig:explore}).}. 

Motivated by group theory, \cite{Agarwal:2023xwl} introduced a new set of variables $(w,\sigma)$ that are related to the traditional cross ratios as
\begin{equation}\label{eq:wsig}
	w \equiv \frac{1-\sqrt{v}}{\sqrt{u}};\quad\quad \sigma \equiv \frac{1+\sqrt{v}}{\sqrt{u}}.  
\end{equation}
This maps the first quadrant of $\sqrt u$-$\sqrt v$ plane, Fig. \ref{fig:uv}(b), into various regions in  Fig. \ref{fig:uv}(c).  These variables have a natural physical meaning in terms of the radial quantization scheme in the Euclidean region where the four points are arranged in an antipodal configuration on circles \cite{Hogervorst:2013sma}. In terms of the more natural $(\rho,\bar\rho)$ coordinates traditionally used in this setup, it can be shown that
\begin{equation}
	w = \cos\theta = \frac{1}{2}\sqrt{\frac{\bar\rho}{\rho}} + \frac{1}{2}\sqrt{\frac{\rho}{\bar\rho}};\quad \sigma = \cosh\eta = \frac{1}{2}\sqrt{\bar\rho\rho} + \frac{1}{2}\sqrt{\frac{1}{\rho\bar\rho}}\ .
\end{equation}

In \cite{Agarwal:2023xwl}, it was shown that, the region $ \sigma>1 $ is divided into 5 parts (see Fig. \ref{fig:uv}(c)). The Euclidean region E corresponds to $-1< w<1$.   The other 4 are associated with Minkowski signature, now with $1 < |w|$.   We shall focus on the two regions on the right, $M_s$ and $M_s^{(t)}$, i.e., with $1<w$. We have showcased in \cite{Agarwal:2023xwl}  that these two regions are causally distinct and the semigroup description of causality applies here~\footnote{Note that the Wick rotation taking us to the region $\sigma<1$ is also a continuation into Lorentzian region (see Fig. \ref{fig:explore}). This leads to a description of the fourpoint function in terms of discrete series in both $\Delta$ and $\ell$, which is not directly useful in our analysis, where we want to study analyticity in these variables.}. 

The restriction $ 1<\sigma < w$ helps maintain $x_1,x_4$ and $x_3,x_2$ timelike separated, while holding all other separations spacelike. This restricts our problem and gives us a clean look at region $M_s^{(t)}$ where only the double discontinuity $\braket{[\varphi(x_4),\varphi(x_1)][\varphi(x_2),\varphi(x_3)]}$ is non-vanishing, which is the object of interest in the Lorentzian inversion formula.

In the $(\sqrt{u},\sqrt{v})$ plane, this region is located in the fourth quadrant since $\sqrt{v}<0$. It is labelled by $M_s^{(t)}$ in Fig. \ref{fig:uv}(c) and Fig. \ref{fig:explore}. There are two OPE limits one can take in this region: $x^2_{12},x^2_{34}\rightarrow0^-$ (which corresponds to the point $T'$), or $x^2_{14},x^2_{23}\rightarrow0^+$ (which corresponds to the point $U$~\footnote{Note that this limit point $U$ should not be confused with the standard Euclidean $u$-channel OPE limit. To approach $U$ from $M_s^{(t)}$, one needs to maintain $\sqrt{v}<0$ with $\sqrt{u}>0$. Physically, this corresponds to $x_4$ approaching $x_1$ within the lightcone and similarly for $x_3$ and $x_2$. In the $(w,\sigma)$ plane, this point corresponds to $(1,1)$ while maintaining $w>\sigma>1$.}). Physically, the point $T'$ corresponds to the forward limit (or Regge limit) that we are interested in. In the $(w,\sigma)$ plane, this limit is located at infinity with $|w|>|\sigma|$ with $w/\sigma \rightarrow \infty$ (see Fig. \ref{fig:uv}(c)).

\begin{figure}[h]
	\centering
	\begin{subfigure}{0.4\textwidth}
		\includegraphics[width=\textwidth]{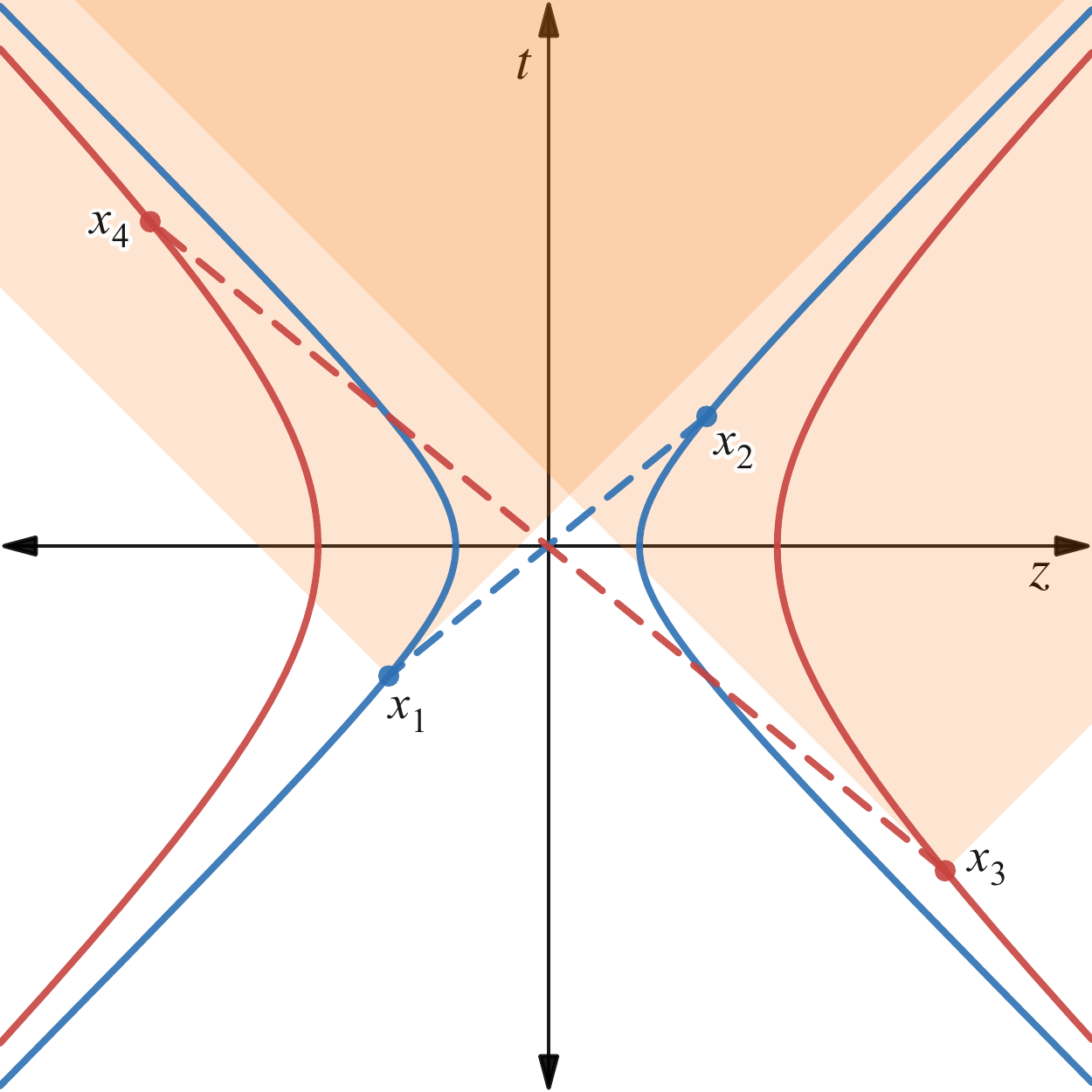}
		\caption{}
	\end{subfigure}
	\caption{Illustration of the $w,\sigma$ variables on the Minkowski spacetime with coordinates $(z,t)$. Dashed lines illustrate causal scattering configurations where $x_{14}^2<0$ and $x_{23}^2<0$. The variable $e^\eta$ gives the radius of the hyperbole whereas $w$ parametrizes the position of the points.}
	\label{fig:antipodal}
\end{figure}

Transition to the Minkowski setting in these variables, (from $|w|<1$ to $1<|w|$),  is equivalent to a Wick rotation in the $w$ variable. Here, without loss of generality, we adopt the convention $\theta\rightarrow iy$ to get to the configuration shown in Fig. \ref{fig:antipodal}, where the parameter $y$ parametrizes the rapidity in the $z$-direction.

It is also useful to to examine the causal issues in a broader perspective by extending into the whole $(\sqrt u, \sqrt v)$ and $(w, \sigma)$ planes, Fig. \ref{fig:explore}, as done in  \cite{Agarwal:2023xwl}.  This is briefly summarized in Appendix \ref{app:morekinematics}. In particular, we find that another  pair of new variables $(q,\bar q)$ can  be convenient in defining Abel/Radon transform for $d=2$. 

\section{Group Theoretic Identification}\label{sec:groups}

One of the key objectives of this study is to understand the Lorentzian OPE inversion formula of \cite{caron2017analyticity} from a representation theoretic perspective. A key observation of \cite{Agarwal:2023xwl}~\footnote{The analogous Euclidean result was first outlined in \cite{Schomerus2017}.} is the fact that, in applying to CFTs, the double discontinuity of the reduced invariant fourpoint scalar amplitudes, $f(u,v) = \text{dDisc}F(u,v)$, when treated as functions over $SO(d,2)$,  are $ H $-bi-invariant,
\be
f(h g h') =f(g),
\ee
with $h,h'\in H$ an appropriate non-compact subgroup. In particular, it was established that, for unitary irreducible representations, the functions ${\cal G}_{(\widetilde{\ell},\widetilde{\Delta};0)}(g)$ are the zonal spherical functions defined over the causal region $M_s^{(t)}$, Fig. \ref{fig:uv}(c). The fact that Minkowski fourpoint functions are distributions has been pointed out in several recent works \cite{Kravchuk:2020scc,Kravchuk:2021kwe,Qiao:2022,Agarwal:2023xwl,Costa:2023wfz}. In fact, causality restricts these functions to be defined piecewise over \textit{causal semigroups}. Explicit examples for $ d=1,2 $ were illustrated in \cite{Agarwal:2023xwl}.

It is crucial to identify the associated space of functions with that of invariant conformal cross ratios, e.g., $(\sqrt u, \sqrt v)$. The $H$-bi-invariant zonal spherical functions, under a Cartan decomposition of the group $g = hah'$, were shown to be functions of a two dimensional abelian subgroup $A(y,\eta)$ comprising of the Lorentz boost and dilatation. It was shown that this group parametrizes the cross ratios by Eq. \ref{eq:wsig}.

\subsection{Construction of $H$-bi-invariant Functions}\label{sec:review}
In \cite{Agarwal:2023xwl}, an $H$-bi-invariant basis of representation functions associated with these semigroups was constructed to serve the purpose of conformal blocks for the region where the double discontinuity is non-vanishing~\footnote{The Minkowski conformal blocks thus constructed were first highlighted in \cite{Raben:2018rbn}}. Without a direct reference to the conformal Casimir equation, one can use the formalism of induced representations to construct these functions (for an introduction to induced representations, see \cite{knapp1986,knapp2002lie}; for application to Lorentz group, see \cite{Bargmann:1947}; for application to $SL(2\mathbb{R})$, see \cite{lang,sugiura}). The central idea of this construction is to write the unknown representation of the full group as a product of known representations of its subgroups.

It becomes particularly useful to write group elements in a Gram-Schmidt like product. Consider the space $\mathbb{R}^n$ for example. A set of $n$ orthogonal basis vectors can be constructed out of any given set of $n$ linearly independent vectors in a procedural way. We start by fixing some vector $\bold{u}_1$ and rescaling it to get a unit vector $\hat{\bold{v}}_1$. A second vector $\bold{u}_2$ can be written as a linear combination of the unit vector $\hat{\bold{v}}_1$ along with a component orthogonal to $\hat{\bold{v}}_1$. That is, $\bold{u}_2 = (\bold{u}_2\cdot \hat{\bold{v}}_1)\hat{\bold{v}}_1 + \bold{v}_2$. Since $\bold{v}_2$ is orthogonal to $\hat{\bold{v}}_1$, the unit vector in its direction gives us our second basis vector $\hat{\bold{v}}_2$, and so on for $\hat{\bold{v}}_3,\hat{\bold{v}}_4,\dots,\hat{\bold{v}}_n$~\footnote{ The $k^{\text{th}}$ vector in the sequence is written as a linear combination of the $k-1$ vectors preceding it, and a component orthogonal to these $k-1$ vectors. In the defining representation for $\mathbb{R}^n$, these can be written in terms of triangular matrices.}.

In general, for a space that has the isometry group $G$, this is equivalent to the Iwasawa decomposition, written as $G = NA_IK$. Here, $N$ is the set of nilpotents that can be represented as triangular matrices (and therefore enables one to form linear combinations of vectors), $A_I$ is the MASG (these help rescale the vectors to unit length), and $K$ is the maximal compact subgroup. In the context of induced representations, a general representation over $G$ can be written as a product of representations over $K$ and $A_I$. Unitary representations over the nilpotents $N$ can be safely mapped to 1 \cite{knapp1986,knapp2002lie,helgason2022groups}.

For our purposes, we note that elements of the semigroup are contained in group elements of the form $NA_IH$ \cite{Agarwal:2023xwl,Hilgert:1996,Faraut:1986}. Since the subgroup $A_I$ is a product of two one-dimensional commuting subgroups, its unitary representations are simply $e^{\widetilde{\ell}y_I}\times e^{-\widetilde{\Delta}\eta_I}$, with $\widetilde{\ell},\widetilde{\Delta}$ imaginary, and the negative sign with $\widetilde{\Delta}$ is a matter of convention. 

Further, since we are looking for $H$-bi-invariant functions, the unitary representations of $H$ that are of interest to us are the ones that transform as identity. Therefore, we start by considering functions over the coset $G/H$. Elements of these cosets have the form $HA$ in the Cartan decomposition and $NA_I$ in the Iwasawa decomposition~\footnote{For clarity of notation, we shall qualify groups and their parameters with a subscript $I$ to indicate when they are associated with the Iwasawa decomposition. The Cartan decomposition is more privileged and goes without a subscript.}. 

In order to find $H$-bi-invariant representations, it becomes useful to compare these two decompositions on the coset. For this purpose, a procedure was introduced in \cite{Agarwal:2023xwl} that related elements on the coset $G/H$ in the Iwasawa decomposition and Cartan decomposition by relating their squares:
\begin{equation}\label{eq:masterh}
	NA_I \cdot P \cdot A_I^TN^T = HA \cdot P \cdot A^T H^T
\end{equation}
where $P$ is such that $H\cdot P\cdot H^T = P$, and the superscript $T$ indicates taking a transpose. As was the case with \cite{Agarwal:2023xwl}, this equation will be of central importance in the current study~\footnote{This definition of coset elements via a projection operator $P$ is a standard way of studying Grassmannian manifolds such as $SO(n)/(SO(k)\times SO(n-k))$ (see for example \cite{bendokat2024grassmann} and references therein). In \cite{Agarwal:2023xwl}, we generalized this to study the conformal Lorentzian Grassmannian $SO(d,2)/(SO(1,1)\times SO(d-1,1))$.}. 

Since the subgroup $A_I$ is two dimensional and parametrized by $y_I,\eta_I$, we can relate the unitary representations in the Iwasawa decomposition to the Cartan decomposition. Following through with this calculation gives:
\begin{equation}\label{eq:characters}
	e^{\lambda_y y_I} = W_1(y,\eta;h;\lambda_y);\quad e^{\lambda_\eta \eta_I} = W_2(y,\eta;h;\lambda_\eta)
\end{equation}
where $h$ is the set of variables that parametrize the subgroup $H$ on the RHS of Eq. \ref{eq:masterh}, and $\lambda_y,\lambda_\eta$ are representation labels. The results for $d=1,2$ cases are summarized in Tab. \ref{tab:results}. $H$-bi-invariant functions can be constructed by integrating $W_1\times W_2$ over $H$:
\begin{equation}\label{eq:sphr}
	\varphi_{\lambda_y,\lambda_\eta}(y,\eta) = \int_H W_1(y,\eta;h;\lambda_y)W_2(y,\eta;h;\lambda_\eta) dh\ .
\end{equation}
This was done for the case of $d=1,2$ in \cite{Agarwal:2023xwl}. These functions solve the diagonalized conformal Casimir equation, which can be reinterpreted as the radial component of the Laplace-Beltrami operator on the manifold $G/H$. For a review of this diagonalization, see App. \ref{app:casimir}.

\begin{table}
\def\arraystretch{1.8}
\centering
\makebox[\linewidth]{
\begin{tabular}{ c|c|c|c|c } 
	\hline
	$\boldsymbol{d}$ & $ \boldsymbol{\vec{a}_I} $ & $ \boldsymbol{\vec{\lambda}} $ & $ \boldsymbol{H} $\textbf{\{params.\}} & \textbf{Characters} $ (\boldsymbol{e^{\vec{a}_I}}) $\\
	\hline \hline
	$1$ & $ \eta_I $ & $ -\frac{1}{2}-\widetilde\Delta $ & $ SO(1,1)\{\alpha\} $ & $ e^{\eta_I} = (\cosh \eta+ \cosh\alpha \sinh \eta) $ \\ \hline
	\multirow{3}{*}{$2$} & \multirow{3}{*}{$(y_I,\eta_I)$} & \multirow{3}{*}{$\left(-1+\widetilde\ell,-\widetilde\Delta\right)$} & \multirow{3}{*}{$SO(1,1)\times SO(1,1)$} & $ e^{y_I} = \Big(\cosh(y+\eta)[1+\cosh(\alpha-\beta)\tanh(y+\eta)] $\\[-8pt]
	& & & & $\times \cosh(y-\eta) [1+\cosh(\alpha+\beta)\tanh(y-\eta)]\Big)^{1/2}$\\
	\cline{5-5}
	& & & $\{\alpha,\beta\}$ & $e^{\eta_I} = \left(\frac{\cosh(y+\eta)}{\cosh(y-\eta)}\frac{[1+\cosh(\alpha-\beta)\tanh(y+\eta)]}{[1+\cosh(\alpha+\beta)\tanh(y-\eta)]}\right)^{1/2}$ \\[2pt] \hline
\end{tabular}}
\caption{Relating variables for the Iwasawa and Cartan decompositions \cite{Agarwal:2023xwl}. Zonal spherical functions are recovered by using Eq. \ref{eq:sphr} with these results.}
\label{tab:results}
\end{table}

\section{Inversion}\label{sec:inv}
In standard texts on representation theory, for example \cite{lang} that covers the case of $SL(2\mathbb{R})$, the problem of inversion is discussed in significant detail. In particular, a diagram of the form Fig. \ref{fig:transforms} is standard. It essentially states that the spherical transform of a function is equivalent to a Radon transform, followed by a Laplace/Mellin transform (see App. \ref{app:disc} for a treatment of H$_2$).

A generic sketch following \cite{lang,sugiura,beerends1987introduction} of how this might be done is as follows. The bi-invariant functions of interest can be defined by Eq. \ref{eq:sphr} by integrating the inductive characters over the subgroup $H$~\footnote{Here, we use $H$ generically. It can be any subgroup obeying the definition of a symmetric space.}:
\begin{equation}
	\phi_\lambda(a) = \int_H e^{\lambda a_I}\ dh
\end{equation}
This defines spherical functions that can in principle be used for inversion by defining spherical transform. Let there be a reduced fourpoint function $f(a)$. This is an $H$-bi-invariant function since it reduces to the function of cross ratios. A spherical transform can be defined as
\begin{align}\label{eq:spherical}
	\hat{f}(\lambda) &= \int_A f(a)\phi^*_\lambda(a)\ da\\
	&= \int_A f(a)\int_H e^{\lambda^* a_I}\ dh\ da
\end{align}
which assumes square integrability of the basis functions $\phi_\lambda(a)$ and is a statement of orthogonality of this basis. Next, we use the Master Equation \ref{eq:masterh} to realize that all coset elements of the form $ha$ can be rewritten as $n_+a_I$~\footnote{The geometry of the coset manifold defines the integration measure and takes a central role in this discussion. This becomes non-trivial when discussing CFTs since the coset is a section of the null cone in $\mathbb{R}^{d,2}$.}. Therefore, an integral over $ha$ can be turned into an integral over $n_+a_I$ with an appropriately defined measure $\alpha$. This leads to:
\begin{align}\label{eq:s=hm}
	\hat{f}(\lambda)&= \int_{N_+} f(n_+a_I)\ dn_+ \int_{A_I} e^{\lambda^* a_I}\alpha\ da_I\ .
\end{align}
We define the Abel/Radon transform
\begin{equation}\label{eq:abel}
	F_f(a_I) = \int_{N_+} f(n_+a_I)\ dn_+\ .
\end{equation}
Therefore,
\begin{equation}\label{eq:mellinF}
	\hat{f}(\lambda)= \int_{A_I} F_f(a_I)\ e^{\lambda^* a_I}\alpha\ da_I\ = \widetilde{F}_f(\lambda)
\end{equation}
where $\widetilde{F}_f$ is the Laplace transform of $F_f$.

This is usually done for functions which can be decomposed in a basis of $K$-bi-invariant functions over the group. As was emphasized in \cite{Agarwal:2023xwl}, this is not the case for Minkowski CFT where causality restricts us to distributions described in terms of $H$-bi-invariant functions \footnote{The rank-2 nature of $SO(d,2)$ introduces further complications.}. For the case where the MASG is of rank 2, the representation labels $\lambda$ will need to be promoted to a 2-vector $\vec{\lambda}$ along with the group variables $a\rightarrow \vec{a}$ (see Tab. \ref{tab:results} and \cite{Agarwal:2023xwl}). In this section, we work with the $d = 1, 2$ cases as examples to establish some concrete results.

\subsection{Examples}\label{sec:examples}
\subsubsection{$d=1$}
The subclass of 1 dimensional CFTs are of great physical significance (see for example \cite{Mazac:2018qmi}), particularly for the conformal limit of models like the SYK model \cite{kitaev1,kitaev2,Maldacena:2016hyu} as well as for studying conformal line defects. It is well known that for $d = 1$, one typically needs to include both the discrete and principal series representations of the conformal group in the partial wave expansion of correlation functions. In the course of computing the sum of ladder diagrams, the discrete series ends up cancelling the spurious pole contributions coming from the principal series. We will show that this entire phenomena occurs quite naturally in our description.

\subsubsection{Defining Inversion via Abel + Laplace Transform}
The very first ingredient in this approach to the inversion formula is to define the Abel/Radon transform of the reduced fourpoint function and formalize Eq. \ref{eq:abel}. In practice, it is more convenient rewriting the transform as an integral over the Cartan maximal abelian subgroup $a$, instead of the nilpotents:
\begin{equation}
	F_f(a_I) = \int_{a} f(a)\ \mu(a,a_I)\ da
\end{equation}
which introduces a measure $\mu(a,a_I)$ that depends on the manifold under consideration and can be obtained by a simple change of variables established using Eq. \ref{eq:masterh}. For the $SO(1,2)/SO(1,1)$ case under consideration, this is given by:
\begin{equation}
	dn = \frac{e^{-\eta_I/2}}{\sqrt{2}\left(\cosh\eta_I-\cosh\eta\right)^{1/2}}\ d(\cosh\eta)
\end{equation}
which gives the measure
\begin{equation}\label{eq:radonmeasureh}
	\hat\mu(a,a_I) = \frac{e^{-\eta_I/2}}{\sqrt{2}\left(\cosh\eta_I-\cosh\eta\right)^{1/2}}\ .
\end{equation}
This is well defined in the region $1<\cosh\eta<\cosh\eta_I$ and serves as the domain of integration~\footnote{There are two definitions of Abel/Radon transform in the literature: Eq. \ref{eq:radonmeasureh} or Eq. \ref{eq:radonmeasurek}. For an intuitive discussion of these measures in terms of geodesics on the manifold, see App. \ref{app:disc}.}. Traditionally, the Radon/Abel transform is defined with the measure $\mu = \hat\mu e^{\eta_I/2}$:
\begin{equation}
	F_f(\cosh\eta_I) = \frac{1}{\sqrt{2}}\int_{1}^{\cosh\eta_I} \frac{f(\cosh\eta)}{\sqrt{\cosh\eta_I-\cosh\eta}}\ d(\cosh\eta)\ .
\end{equation}

The partial wave amplitude is therefore given by the Laplace transform:
\begin{align}\label{eq:invlap}
	a(\lambda)\equiv\widetilde{F_f}(\widetilde\lambda) &= \int_{0}^\infty F_f(\cosh\eta_I)\ e^{-\lambda \eta_I}e^{-\eta_I/2}\ d\eta_I\nn 
	&= \int_{0}^\infty F_f(\cosh\eta_I)\ e^{-\widetilde\lambda \eta_I}\ d\eta_I
\end{align}
where $\lambda = -1/2 - \widetilde{\Delta}$ and $\lambda^* = -1/2 + \widetilde{\Delta}$. Comparing with Eq. \ref{eq:s=hm}, we find $\alpha = e^{\eta_I/2}$. The complex $\Delta$ plane is identified with the complex $\widetilde\lambda$ plane by $\Delta = 1/2-\widetilde\lambda$. For physical theories, we require $\Delta\geq 0$. Since the spectrum of the theory is captured by the location of singularities of $a(\lambda)$, all physical operators correspond to poles located to the left of the $\widetilde\lambda = 1/2$ line in the complex $\widetilde\lambda$ plane.

\textit{Equivalence with the Inverse Spherical Transform:} We can perform a check to make sure that Eq. \ref{eq:invlap} is indeed equivalent to Eq. \ref{eq:spherical} by swapping the order of integration:
\begin{align}
	a(\lambda)&=\frac{1}{\sqrt{2}}\int_1^\infty  d(\cosh\eta) f(\cosh\eta)\int_{\eta}^\infty{d \eta_I} \frac{e^{(-\lambda-1/2) \eta_I}}{\sqrt{\cosh\eta_I-\cosh\eta}}\nn
	&= \int_1^\infty  d(\cosh\eta) f(\cosh\eta) Q_{-\lambda^*-1}(\cosh\eta)\nn
	&= \int_1^\infty  d(\cosh\eta) f(\cosh\eta) \varphi_{\lambda^*}(\cosh\eta) = \hat{f}(\lambda)
\end{align}
This shows that the partial wave amplitudes recovered from the Lorentzian OPE inversion formula $a(\lambda)$ is equivalent to the Mellin transform of the Radon transformed function $\widetilde{F}_f(\widetilde\lambda)$. As mentioned before, this can be generalized to $d>1$ by promoting the representation labels to 2-vectors.

It is interesting to note that we have recovered an alternate definition of zonal spherical functions (which serve as the principal series analytic continuation of Minkowski conformal blocks) in terms of the Radon transform measure:
\begin{equation}\label{eq:sphr-radon}
	\varphi_{\lambda}(a) = \int_{a_I\geq a} da_I\ \mu(a,a_I)\ e^{-\widetilde\lambda a_I}\ .
\end{equation}

\subsubsection{Recovering the Fourpoint Function}
We can reverse the problem and ask how to reconstruct the fourpoint function (or dDisc thereof) for a given partial wave amplitude $a(\lambda)$. One way to answer this is by inverting the procedure developed so far in this section. We start with the partial wave amplitudes and perform an inverse Laplace transform, followed by an inverse Radon transform. Let us start with the first step:
\begin{equation}
	F_f(\cosh\eta_I) = \int_{1/2-i\infty}^{1/2+i\infty}\frac{d\widetilde\lambda}{2i\pi}\ e^{(-1/2+\widetilde\lambda) \eta_I}\ a(-1/2+\widetilde\lambda)\ .
\end{equation}
For $\eta>0$, it is possible to close the contour of this integral to the left and pick up the residues corresponding to all the relevant physical singularities. Further, if we assume that the partial wave amplitudes $a(\lambda)$ are analytic to the right (we argued that all physical singularities lie to the left in the previous section), we can also conclude that 
\begin{equation}\label{eq:ranalytic}
	\int_{1/2-i\infty}^{1/2+i\infty}\frac{d\widetilde\lambda}{2i\pi}\ e^{(1/2-\widetilde\lambda) \eta_I}\ a(-1/2+\widetilde\lambda) = 0\ .
\end{equation}

We can now define the inverse Radon transform (there are several definitions that are equivalent \cite{Helgason_1999,samko1993fractional}):
\begin{equation}
	f(\cosh\eta)= \frac{1}{\pi\sqrt{2}}\int_{1}^{\cosh\eta} d(\cosh\eta_I)\frac{F_f'(\cosh\eta_I)}{\sqrt {\cosh\eta-\cosh\eta_I}}  
\end{equation}
where
\begin{equation}
	F_f'(\cosh\eta_I) = \frac{d\ F_f(\cosh\eta_I)}{d(\cosh\eta_I)}\ .
\end{equation}
This leads to a new representation of the function:
\begin{align}
	f(\cosh\eta)&= \frac{1}{\pi\sqrt{2}}  \int^{\eta}_0 \frac{d\eta_I}{\sqrt{\cosh\eta-\cosh\eta_I}}\int_{1/2-i\infty}^{1/2+i\infty}\frac{d\widetilde\lambda}{2i\pi } (-1/2+\widetilde\lambda) e^{(-1/2+\widetilde\lambda )\eta_I} a(-1/2+\widetilde\lambda)  \\
	&= \frac{1}{\pi\sqrt{2}} \int_{1/2-i\infty}^{1/2+i\infty}\frac{d\widetilde\lambda}{2i\pi } (-1/2+\widetilde\lambda)a(-1/2+\widetilde\lambda)\int^{\eta}_0 d\eta_I\frac{e^{(-1/2+\widetilde\lambda )\eta_I}}{\sqrt{\cosh\eta-\cosh\eta_I}}
\end{align}

As is, the integral over $\eta_I$ does not have a closed form. However, since the amplitudes are analytic to the right, one can add a term similar to Eq. \ref{eq:ranalytic} without loss of generality to get:
\begin{align}\label{eq:pwd}
	f(\cosh\eta) &= \frac{2}{\pi\sqrt{2}} \int_{1/2-i\infty}^{1/2+i\infty}\frac{d\widetilde\lambda}{2i\pi } (-1/2+\widetilde\lambda) a(-1/2+\widetilde\lambda)\int^{\eta}_{0} d\eta_I\frac{\cosh((-1/2+\widetilde\lambda )\eta_I)}{\sqrt{\cosh\eta-\cosh\eta_I}}\nn
	&= \int_{1/2-i\infty}^{1/2+i\infty}\frac{d\widetilde\lambda}{2i\pi } (-1/2+\widetilde\lambda)a(-1/2+\widetilde\lambda) P_{\widetilde\lambda-1}(\cosh\eta)
\end{align}
where $P_{\widetilde\lambda-1}(\cosh\eta)$ are Legendre functions of the first kind. If the zonal spherical functions $Q_{\lambda}(\cosh\eta)$ (which are Legendre functions of second kind) are thought of as the conformal blocks, then the above decomposition can be thought of as the partial wave decomposition of the function. Clearly, the partial waves $P$ can be written as a linear combination of the block and shadow block.

\textit{Changing Basis:} Let us make this change of basis to conformal blocks more explicit by use of the identity
\begin{equation}
	P_\nu(z) =\frac{\tan \nu \pi}{\pi}\Big( Q_\nu(z)-Q_{-\nu-1}(z)\Big)\ .
\end{equation}
After a shift by 1/2, this leads to:
\begin{align}
	f(\cosh\eta) &= -\int_{-i\infty}^{+i\infty}\frac{d\widetilde\lambda}{2i\pi } (\widetilde\lambda)a(\widetilde\lambda) \frac{\cot \widetilde\lambda \pi}{\pi}\Big( Q_{\widetilde\lambda-1/2}(\cosh\eta)-Q_{-\widetilde\lambda-1/2}(\cosh\eta)\Big)\nn
	&= \int_{-i\infty}^{+i\infty}\frac{d\widetilde\lambda}{2i\pi } (\widetilde\lambda)\left(a(\widetilde\lambda)-a(-\widetilde\lambda)\right) \frac{\cot \widetilde\lambda \pi}{\pi} Q_{-\widetilde\lambda-1/2}(\cosh\eta)
\end{align}
When the contour is closed in the above representation, one recovers the conformal block expansion of the fourpoint function. Note that this case is completely equivalent to the approach taken in \cite{Maldacena:2016hyu} for the conformal limit of the SYK model~\footnote{This can be seen by relating to their variable $\chi = \sqrt{u} = 2/(1+\cosh\eta)$, writing Legendre $Q$ functions in terms of hypergeometric functions $Q_\nu(z) = \sqrt{\pi}\frac{\Gamma(\nu+1)}{\Gamma(\nu+3/2)}(2z)^{-(\nu+1)} F(\nu/2+1/2,\nu/2+1;\nu+3/2,1/z^2)$, and using the identity $F(a,b;2b,z) = (1-z/2)^{-a}F(a/2,a/2+1/2;b+1/2,z^2/(2-z)^2)$.} where they include both the principal series and the discrete series when calculating the sum of ladder diagrams. The discrete series is recovered from the $a(-\widetilde{\lambda})\cot\widetilde{\lambda}\pi$ term by closing the contour to the left where $a(-\widetilde{\lambda})$ is analytic.

\subsection{$d=2$}
The group theoretic discussion in Sec. \ref{sec:inv} is fairly general. However, solving the Master Eq. \ref{eq:masterh} for higher dimensional cases to get the right measures for the transform gets tedious. We will illustrate the $d = 2$ case here, in which the isomorphism $SO(2,2)\cong SL(2\mathbb{R})\times SL(2\mathbb{R})$ means that the result reduces to a product of two copies of the $d = 1$ case discussed earlier. This case brings out several subtleties about the discrete symmetries of the partial wave amplitudes that will be of interest to us.

\subsubsection{Defining the Abel/Radon Transform}
We can define the Abel/Radon transform similarly as before from Eq. \ref{eq:abel} to get our auxiliary fourpoint function. The integral over the nilpotents is defined over the set of positive roots of the MASG $A$. For the $d=2$ case in the region $M_s^{(t)}$, these correspond to the null translations generated by $T_t + T_z$ and the special conformal transformation $C_t - C_z$ \cite{Agarwal:2023xwl}. In the diagonal basis, these form the set of lower triangular matrices.

From the Master Equation \ref{eq:masterh}, we can change variables from $t_+, c_-$ to $\cosh(y-\eta), \cosh(y+\eta)$. This gives:
\begin{align}
	dt_+ &= \frac{1}{2\sqrt{2}}\frac{e^{(-y_I+\eta_I)/2}\ d(\cosh(y-\eta))}{\left(\cosh(y_I-\eta_I) - \cosh(y-\eta)\right)^{1/2}}\\
	dc_- &= \frac{1}{2\sqrt{2}}\frac{e^{(-y_I-\eta_I)/2}\ d(\cosh(y+\eta))}{\left(\cosh(y_I+\eta_I) - \cosh(y+\eta)\right)^{1/2}}
\end{align}

\subsubsection{An Example: Generalized Free-Fields for $d=2$ in terms of $(w,\sigma)$}
For the mean field example, we have:
\begin{align}
	\text{dDisc}\ {\cal G} &= \frac{2^{1+2\Delta}\sin^2(\Delta\pi)}{(\cosh y - \cosh\eta)^{2\Delta}}\nn 
	&= \frac{2^{1+2\Delta}\sin^2(\Delta\pi)}{((\cosh(y-\eta)-1) (\cosh(y+\eta)-1))^{\Delta}}
\end{align}

The Radon transform factorizes and is given by:
\begin{align}
	{\cal F}_{\Delta,\ell}(y_I,\eta_I) &= \int_{n_+} \text{dDisc}\ {\cal G}(a_In_+)\ dn_+\nn
	&=\ \frac{2^{1+2\Delta}\sin^2(\Delta\pi)}{8}\left(\frac{\Gamma(1-\Delta)}{\Gamma(3/2-\Delta)}\right)^2
	(\cosh(y_I - \eta_I)-1)^{1/2-\Delta}(\cosh(y_I + \eta_I)-1)^{1/2-\Delta}
\end{align}
The above result applies only if $\mathbb{R}(\Delta)<1$.

To extract the partial wave amplitudes from this, we need to perform a Laplace transform:
\begin{align}
	\hat{f}(\widetilde\Delta,\widetilde\ell) &= \int_{0}^{\infty}dy_I\int_{-y_I}^{y_I}d\eta_I {\cal F}_{\Delta,\ell}(y_I,\eta_I) e^{-\widetilde{\ell}y_I}e^{\widetilde{\Delta}\eta_I}\nn
	&= 2\int_{0}^{\infty}d(y_I+\eta_I)\int_{0}^{\infty}d(y_I-\eta_I) {\cal F}_{\Delta,\ell}(y_I,\eta_I) e^{(-\widetilde{\ell}+\widetilde{\Delta})(y_I+\eta_I)/2}e^{(-\widetilde{\ell}-\widetilde{\Delta})(y_I-\eta_I)/2}
\end{align}
Define:
\begin{equation}
	a = \frac{(1+\widetilde{\ell}-\widetilde{\Delta})}{2};\quad 
	a' = \frac{(1+\widetilde{\ell}+\widetilde{\Delta})}{2}
\end{equation}
This gives the final result:
\begin{equation}\label{eq:ampmf}
	\hat{f}(\widetilde\Delta,\widetilde\ell) = \frac{\pi\Gamma(1-\Delta)^2 }{\Gamma(\Delta)^2}\frac{\Gamma(a+\Delta-1)}{\Gamma(a-\Delta+1)}\frac{\Gamma(a'+\Delta-1)}{\Gamma(a'-\Delta+1)}\ .
\end{equation}
The generalized free field example is known to be closely related to the SYK model as it can serve as the kernel for the construction of ladder diagrams via an equation of Volterra type (see chapter 8 of \cite{Hilgert:1996} for a general treatment of Volterra algebras and their connection with harmonic analysis). Our result of Eq. \ref{eq:ampmf} is related to the bosonic version of 2-dimensional SYK model discussed in \cite{Murugan:2017eto} by the identification $a = h$ and $1-a' = \widetilde{h}$. Note that this implies that their amplitude was symmetric in $\widetilde{\ell}$, which is to be expected for a Euclidean treatment (which is continued into $M_s$). Since our treatment focusses on the region $M_s^{(t)}$, where the double discontinuity lives, the partial waves are symmetric in $\widetilde{\Delta}$ instead.

\subsubsection{A Discussion on Symmetries of the Partial Wave Amplitudes}
Note that Eq. \ref{eq:ampmf} is symmetric in $\widetilde\Delta$. This is reflective of the fact that the inversion is symmetric in $\eta_I$. Note that the wedge $M_s^{(t)}$ in Fig. \ref{fig:uv}(c) is still double sheeted in the sense that it only restricts $w>\sigma$ whereas the semigroup restriction is meant to be $0<|\eta|<y$ \cite{Agarwal:2023xwl}. This is best shown by going to the $(q,\bar{q}) = (\cosh(y+\eta),\cosh(y-\eta))$ variables discussed in \cite{Raben:2018rbn}~\footnote{These are related to the traditional $(z,\bar{z})$ by $q = 2/z-1$.}. Here, we treat them as independent variables with $-y<\eta< y$ and $0<y<\infty$. This takes us from the wedge $1<\sigma<w<\infty$ to the square $1<q,\bar{q}<\infty$. 

\begin{table}[ht]
	\centering
	\begin{tabular}{ c|c } 
		\hline
		\textbf{Roots} & \textbf{Generators} \\
		\hline \hline
		$(1,-1);(-1,1)$ & $T_t+T_z;C_t+C_z$ \\ \hline
		$(1,1);(-1,-1)$ & $C_t-C_z;T_t-T_z$ \\ \hline
	\end{tabular}
	\caption{Roots for $SO(2,2)$ ordered as $(L,D)$. Positive roots that belong in $N^+$ are therefore $\{T_t+T_z,C_t-C_z\}$ (see \cite{Agarwal:2023xwl}).}
	\label{tab:roots}
\end{table}

For $2d$ physics, this is related directly to the root space structure of the conformal group. For $SO(2,2)$ and the Cartan subalgebra defined by the Lorentz boost $L$ and dilatation $D$, the algebra contains 4 nilpotents listed in Tab. \ref{tab:roots}. For our causal semigroup, the roots are ordered as $(L,D)$. As can be read off of Tab. \ref{tab:roots}, the transformation $D\leftrightarrow -D$ leaves the set of nilpotents unchanged. This directly implies that our semigroup is invariant as well under this transformation. It is therefore not surprising that our representation for partial wave amplitudes carries this symmetry.

\begin{figure}[ht]
	\centering
	\begin{subfigure}{0.3\textwidth}
		\includegraphics[width=\textwidth]{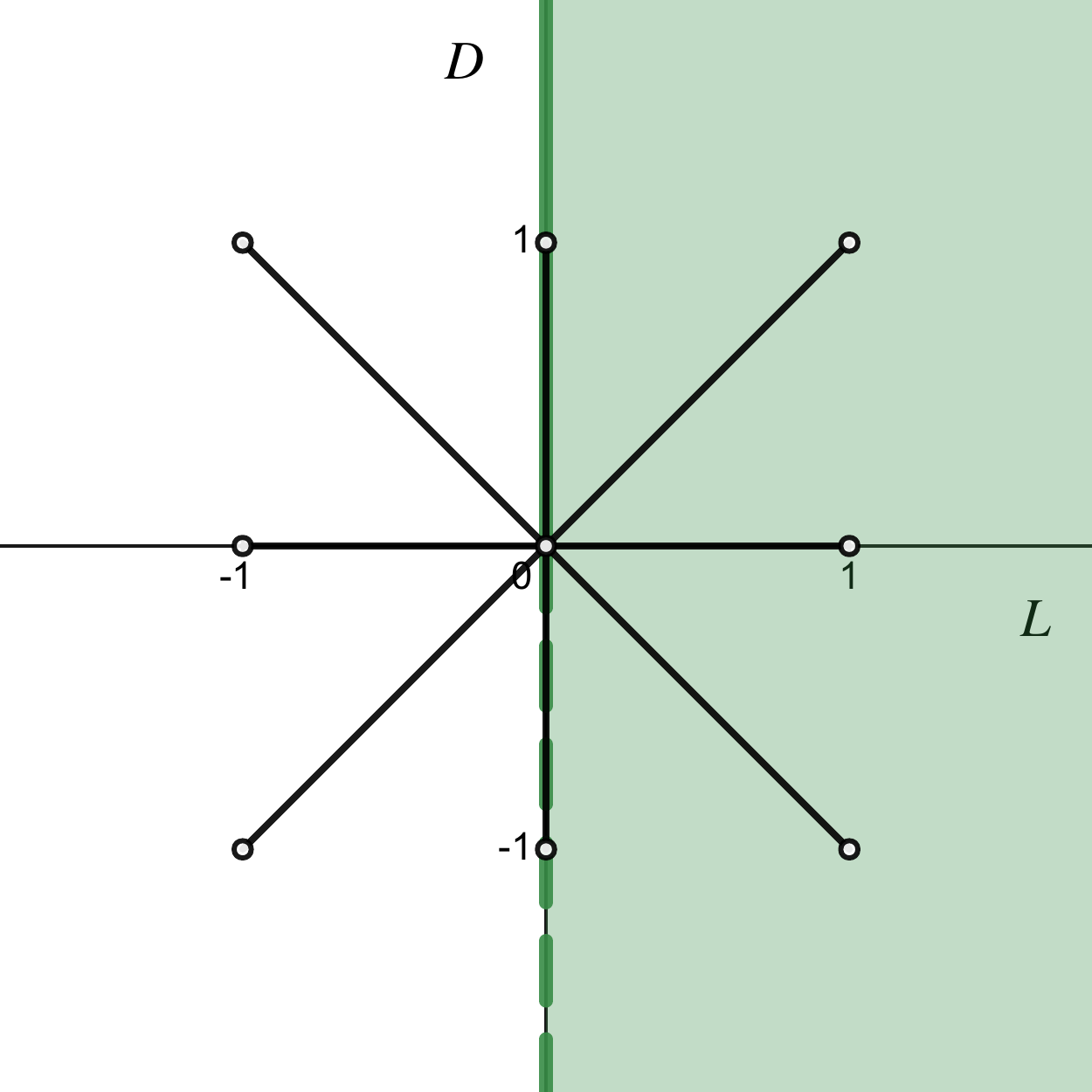}
		\caption{}
	\end{subfigure}
	\quad\quad\quad\quad\quad\quad
	\begin{subfigure}{0.3\textwidth}
		\includegraphics[width=\textwidth]{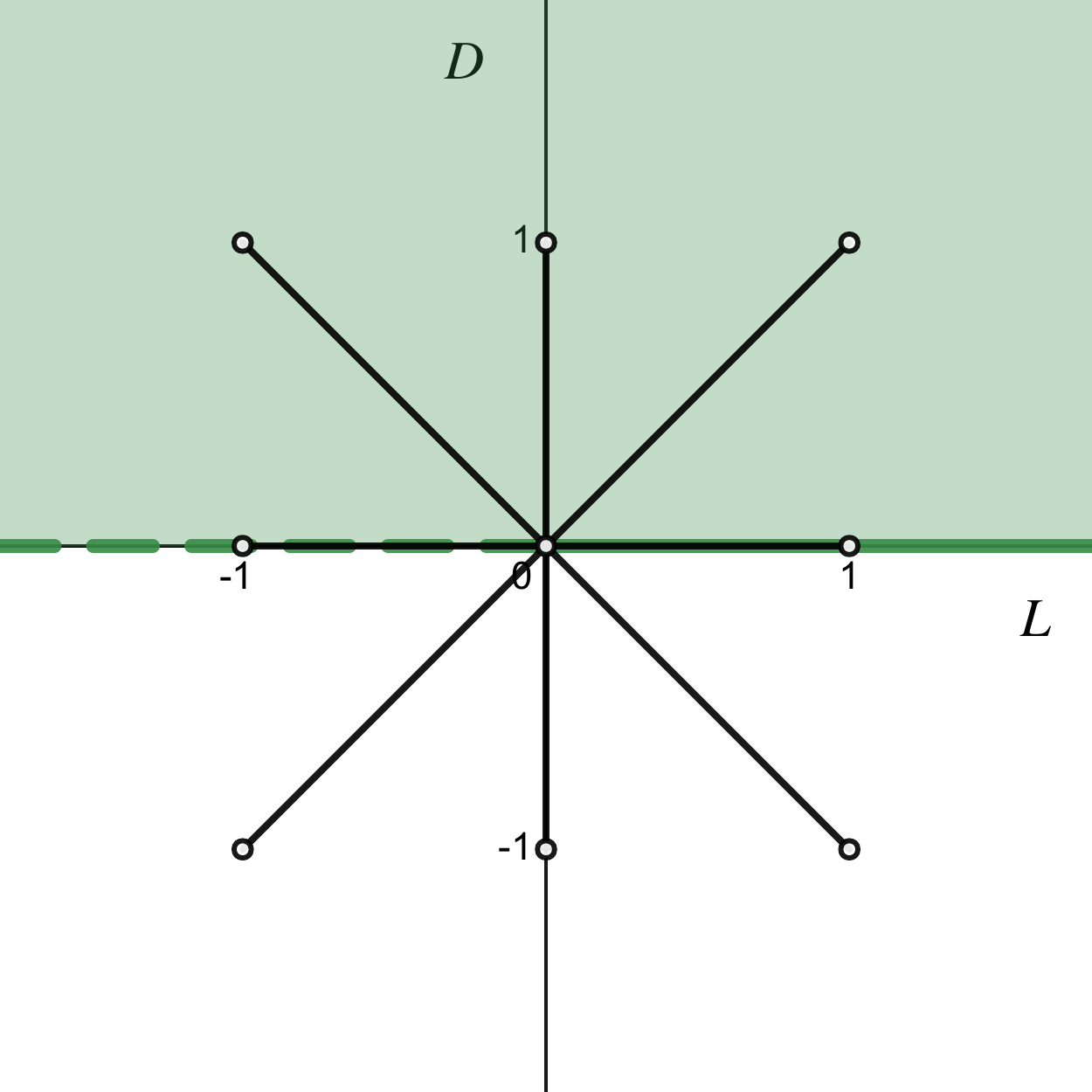}
		\caption{}
	\end{subfigure}
	\caption{The restricted root space diagram for $ SO(d,2) $ is of type $B_2$. The green shaded region contains the set of positive roots. Figure (a) corresponds to ordering $ (L,D) $ whereas figure (b) corresponds to $ (D,L) $. The set of positive roots do not possess any discrete or reflection symmetries.}
	\label{fig:roots}
\end{figure}

However, this is no longer the case for $d>2$. There are in general $2d-2$ roots corresponding to 8 set of positive roots that do not possess any discrete symmetries (see Fig. \ref{fig:roots}), and so it is reasonable to expect that the relevant semigroup representations do not contain these symmetries for the Lorentzian regions. In particular, for the semigroup, these correspond to the 8 distinct regions given by $0<\eta<y,0<-\eta<y,0<\eta<-y,0<-\eta<-y,0<y<\eta,0<-y<\eta,0<y<-\eta,0<-y<-\eta$. With the symmetry under $\eta\leftrightarrow -\eta$ for the $d = 2$ case, this reduces to the 4 cases discussed in this paper.

\section{Discussion}\label{sec:conclusion}
Our main result is an alternative take on the Lorentzian inversion formula, and is illustrated in Fig. \ref{fig:transforms}. The punchline really is that the Mellin amplitudes associated with the horocyclic Radon transform of the reduced fourpoint function are equivalent to the partial wave amplitudes. We presented a proof of concept by working out the $d = 1,2$ cases.

\textit{Future Directions:} A generalization to higher spacetime dimensions is the next obvious step. Although the construction of $H$-bi-invariant function would work largely as outlined, the handling of nilpotents is expected to be a little more subtle due to the semigroup restrictions. This also has to do with the root space structure of the group. At the level of $d = 2$, the root system of the Lorentzian conformal group is $D_2$ which is isomorphic to $A_1\times A_1$. This factorization is responsible for simplifying the discussion for $d=2$. As discussed, this also plays a role in the discrete symmetries of the partial wave amplitudes. For the $d>2$ case, the root system is $B_2$ (isomorphic to $C_2$ when we go from $(w,\sigma)$ to $(q,\bar{q})$), and selecting a specific semigroup is far more restrictive. It would be interesting to see how this plays a role when constructing partial waves in this view. Intuitively, constructing totally symmetric partial waves would require 8 terms, each corresponding to a unique solution for the Casimir equation with an associated Weyl chamber. How exactly this translates to tomography will be interesting to explore.

The problem of constructing the full Euclidean fourpoint function via dispersion relations using techniques presented in this paper is also interesting. It requires defining the inverse Mellin and inverse Radon transforms appropriate for the Euclidean setting (or more generally for the first quadrant of the $(\sqrt{u},\sqrt{v})$ plane). This leads to a double dispersion relation, similar to \cite{carmi2020conformal,caron2021dispersive}. Although this can be done for specific simple examples, a completely general description requires careful handling of the double cut at $\sqrt{v} = 0$. We leave this problem to future work.

Since the Mellin amplitudes for our auxiliary functions are dependent only on primaries, it would be interesting to see if the bootstrap for these auxiliary functions simplifies. In other words, how the Radon transform behaves under crossing might shed light on the structure of CFTs. The integral representation of conformal blocks given in Eq. \ref{eq:sphr-radon} is completely general and can be used to study their structure. Further, in our approach, the Radon transform is explicitly defined only on the null cone, with no reference to the bulk. It would be interesting to relate our work with that of \cite{Czech:2016xec,Bhowmick:2017uci,Das:2017wae} where Radon transforms were shown to play a role in AdS bulk reconstruction.

Finally, the advocated shift in perspective from conformal Casimir equation to a diagonalized Laplace-Beltrami operator will also be interesting to adapt to the study of higher point functions that has gained recent interest \cite{Antunes:2021kmm,Buric:2020dyz,Buric:2021ywo,Buric:2021ttm,Buric:2021kgy}, particularly for Regge physics \cite{Costa:2023wfz}.

\paragraph{Acknowledgements:} The work of R. C. B. is supported by the U.S. Department of Energy (DOE) under Grant No. DE-SC0015845 and DE-SC0019139. Centro de Física do Porto is partially funded by the Foundation for Science and Technology of Portugal (FCT).

\newpage
\appendix
\addtocontents{toc}{\protect\setcounter{tocdepth}{1}}
\section{More Kinematics}\label{app:morekinematics}
\begin{figure}[h]
	\centering
	\begin{subfigure}{0.45\textwidth}
		\includegraphics[width=\textwidth]{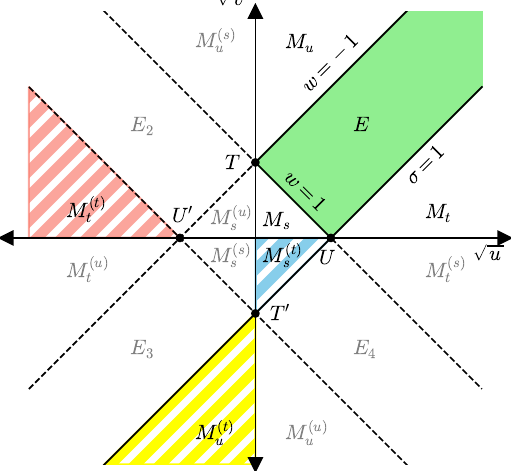}
		\caption{}
	\end{subfigure}
	\quad\quad\quad
	\begin{subfigure}{0.45\textwidth}
		\includegraphics[width=\textwidth]{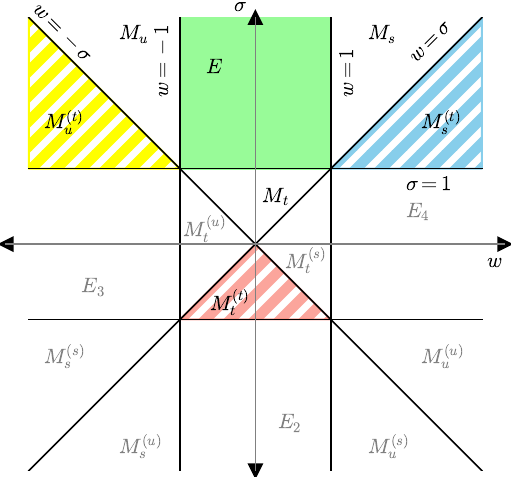}
		\caption{}
	\end{subfigure}
	\caption{ Extension into causal regions in $ \sqrt{u},\sqrt{v} $ plane.  In terms of new variables $(w,\sigma)$, the corresponding 16 regions are shown in Fig. \ref{fig:explore}(b). Special causal regions $M_s^{(t)}$, $M_u^{(t)}$ $M_t^{(t)}$ are shaded.}
	\label{fig:explore}
\end{figure}

Kinematically, it is useful to extend into the whole $(\sqrt u, \sqrt v)$ and $(w, \sigma)$ planes, Fig. \ref{fig:explore}.  Topologically,  Fig. \ref{fig:explore}(b) should be read as a projective plane with a circle at infinity and antipodal points identified. It is tiled by 16 triangular regions and each region contains three limiting points. Typically, when the Lorentzian limit is considered, if the points are held spacelike, the fourpoint function lives in the region $M_s$. Here, in addition to the OPE points $T$ and $U$, we gain access to a particularly interesting limit which is the origin $(\sqrt{u},\sqrt{v})\rightarrow (0^+,0^+)$. This corresponds to the limit $(z,\bar{z})\rightarrow(0^+,1^-)$ and is known in the literature as the double-lightcone limit. This is the limit of interest for the lightcone bootstrap program \cite{Fitzpatrick:2012yx,Komargodski2013,Kaviraj:2015cxa}. In the $(w,\sigma)$ variables, this corresponds to the limit $(w,\sigma)\rightarrow(\infty,\infty)$ with $w/\sigma\rightarrow1^-$.

There is, however, an alternate approach to this limit such that $(z,\bar{z})\rightarrow(1^-,0^+)$. Physically, in this limit the point $x_2$ approaches the past lightcones on points $x_1,x_3$ instead of the future lightcones. These two configurations are equivalent since the operators are commuting. It unfurls the causal triangle $0<z<\bar{z}<1$ to the causal diamond $0<z,\bar{z}<1$ where the fourpoint function is well defined. This discrete symmetry is better understood in terms of another set of variables,
\be
(q,\bar{q})\equiv\left(\frac{2-z}{z},\frac{2-\bar{z}}{\bar{z}}\right)=(\cosh(y+\eta),\cosh(y-\eta)),
\ee
particularly in the context of the region $M_s^{(t)}$.

It can be shown that the lightcone singularities place further restrictions such that $y>0$. In particular, $y>\eta>0\  (\sim q<\bar{q})$ restricts us to the forward lightcone whereas $y>-\eta>0\ (\sim \bar{q}<q)$ restricts us to the backward lightcone. In general, one should consider the entire region $-y<\eta<y$, which is covers the square $1<q,\bar{q}<\infty$. The discrete symmetry in $\eta\leftrightarrow-\eta$ corresponds to the symmetry of partial waves under the scale shadow transform~\footnote{The restriction $w>\sigma$ is more general. One can also consider the branch $y<0$ such that $y<\eta<-y$. This takes us to the past lightcone (instead of the future lightcone case considered in the main text). The entire discussion here repeats, with the subtle difference that one gets representation functions that differ from the ones in discussed by a spin shadow transform. The semigroup restriction allows for either one or the other.}. 

As discussed in \cite{Agarwal:2023xwl}, these discrete symmetries are closely related to the root space structure of the conformal group.

\section{Casimir Equation}\label{app:casimir}
The Casimir Equation for identical scalars is given by:
\begin{align}
	{\cal C}=& - 2[(1-u-v)\partial_{v}(v\partial_{v})+u\partial_{u}(u\partial_{u}-d)\nn
	&-(1+u-v)(u\partial_{u}+v\partial_{v})(u\partial_{u}+v\partial_{v})], \label{eq:Casimir}
\end{align}
with eigenvalues
\begin{equation}
	\nu \equiv C(\ell,\Delta) = -[\Delta (\Delta-d) + \ell (\ell +d-2) ]
	\label{eq:Casimir-value}
\end{equation}
where $\Delta$ real and $\ell$ (non-negative) integers \cite{Dolan:2003hv,Dolan:2000ut,dolan2011conformal}.

Since Eq. \ref{eq:Casimir} follows from the algebra of $SO(d+1,1)$, its realisation for $SO(d,2)$ can be directly obtained by expressing $u$ and $v$ in terms of $SO(d,2)$ group parameters and  its action is now carried out in a Minkowski setting. The Casimir takes on the form
\begin{equation}
	\mathcal{C}=-\partial_{y}^{2}-\partial_{\eta}^{2}-\alpha(y,\eta)\partial_{y}-\beta(y,\eta)\partial_{\eta}  \label{eq:CAB}
\end{equation}
where boost  and dilatation are realized as differential operators, with  $\alpha(y,\eta) $ and $\beta(y,\eta) $ given by
\begin{align}
	\alpha(y,\eta)  &=d-4t^{-2}\left(\frac{1-2\sigma^{2}+t^{-2}}{1+t^{-4}-2t^{-2}(2\sigma^{2}-1)}-\frac{d-2}{2(1-t^{-2})}\right),\\
	\beta(y,\eta)  &=(d-2)\coth\eta+4t^{-2}\coth\eta \left(\frac{2(1-\sigma^{2})}{1+t^{-4}-2t^{-2}(2\sigma^{2}-1)}\right).\label{eq:AB}
\end{align}
where with $\epsilon=(d-2)/2$ and $t=e^y$.
The desired $SO(d,2)$ conformal blocks, ${ G}^M_{\ell, \Delta} (y,\eta)$,   are solutions to 
\begin{equation}
	\mathcal{C}\,\,{ G}^M_{ \ell, \Delta} (y,\eta)=\nu\,  { G}^M_{ \ell, \Delta} (y,\eta)
\end{equation}
with appropriate boundary conditions. As will be shown later, the associated eigenvalue, $\nu $, can be  re-expressed in terms of principal series representation labels, $\widetilde \ell$ and $\widetilde \Delta$.

\subsection{Hermitian Casimir}
To proceed further, it is useful to bring the Casimir into a Hermitian form \cite{agarwal2022application}, i.e., via $\widetilde {\cal C}= |J|^{-1} {\cal C} |J|$, where
\begin{equation}
	\widetilde {\cal C}= -\partial_y^2 -\partial_\eta^2 + V(y,\eta).\label{eq:HCasimir}
\end{equation}
Following through, we can find a Jacobian which, for general $d$, satisfies the following conditions:  
\begin{align}
	\alpha(y,\eta)&=-\frac{2}{J}\partial_{y}J=-2\partial_{y}\log J;\\ \beta(y,\eta)&=-\frac{2}{J}\partial_{\eta}J=-2\partial_{\eta}\log J\ .
\end{align}
By integrating with respect to $y$ first, it leads to
\begin{align}
	\log J & =\log\left[\frac{(\sigma^{2}-1)^{-\epsilon/2}}{(1+t^{4}-2t^{2}(2\sigma^{2}-1))^{1/2}}\right]+\text{const.}(t)\nonumber \\
	J & =\left[\frac{(\sigma^{2}-1)^{-\epsilon}}{(1+t^{4}-2t^{2}(2\sigma^{2}-1))}\right]^{\frac{1}{2}}\times\text{const.}(t).
\end{align}
The unknown function $\text{const.}(t)$ can be fixed by the second equation, leading to, upto a constant,
\begin{align}\label{eq:jacobian}
	J&=(w^{2}-\sigma^{2})^{-\frac{1}{2}} ((w^{2}-1)(\sigma^{2}-1))^{-(d-2)/4}\nonumber\\
	&= \left[\sinh(y+\eta)\sinh(y-\eta)\right]^{-1/2}\left[\sinh y \sinh\eta\right]^{-(d-2)/2}
\end{align}
with $w=\cosh y$ and $\sigma=\cosh \eta$. This  leads to a ``potential" function $V(y,\eta)$ which, after a short calculation, can be shown to take on a simple form,
\begin{equation}
	V = - \frac{1}{\sinh^2 (y+\eta)}-\frac{1}{\sinh^2 (y-\eta)} + \frac{ (d-4)(d-2)}{2} \left[\frac{1}{\sinh^2y}+\frac{1}{\sinh^2\eta }\right]+V_0, \label{eq:pot}
\end{equation}
where $V_0=(d/2)^2+(d/2-1)^2$.

Note immediately that the Jacobian partitions the phase space into regions boundaried by lines $J^{-1}=0$ (labelled by $L_u, L_s$ in Fig. \ref{fig:uv}(c)), with $V(y,\eta)$ singular on the boundaries (see Eq. \ref{eq:pot}). These partitions can be identified with various regions indicated  in Fig. \ref{fig:uv} and \ref{fig:explore}.

In particular, the $ |w|=|\sigma| $ lines dictate the boundary of regions where particular double discontinuities are non-vanishing. For example, the region $\infty>w>\sigma>1$ is the region where
\begin{equation}\label{eq:scausal}
	\braket{[\varphi(x_4),\varphi(x_1)][\varphi(x_2),\varphi(x_3)]}
\end{equation}
is non-zero. Kinematically, this is the region illustrated in Fig. \ref{fig:antipodal}(b) where $x_{14},x_{23}$ are timelike, whereas all other separations are spacelike. In subsequent sections, we will focus on the imaginary parts of the fourpoint function such that they are non-vanishing only in this specific region.

Further, the $ |w|=1$ and $|\sigma| = 1 $ lines, in view of their interpretation as $ w = \cosh y,\ \sigma = \cosh\eta $, indicate a Wick rotation when going across them. For example, starting in region $ M_s $, we can perform a Wick rotation taking the group $ SO(d,2)\rightarrow SO(d+1,1) $ and $ w\rightarrow\cos\theta $, thus moving into the Euclidean region $E$. By the same token, moving to the region $ M_t $ from $E$ requires we take $ \sigma\rightarrow\cos\theta' $ leading to representations of $ SO(d+2) $~\footnote{In Fig. \ref{fig:uv}(c), we have made a point to only draw segments of these lines to indicate regions of immediate relevance. These lines extend beyond and divide the planes into 16 regions. See Fig. \ref{fig:explore}.}.

This form of the Casimir equation \ref{eq:HCasimir} is related to Calogero-Sutherland models and has recently been studied extensively by Schomerus et al. \cite{Schomerus2017,Isachenkov2018,Schomerus2018,Buric2020,Buric:2022ucg}. The potential \ref{eq:pot} itself is insensitive to the causal structure described here in the sense that it carries the same structure in both regions $M_s$ and $M_s^{(t)}$. One can therefore study it in the region $M_s$ (which is still necessarily described by $SO(d,2)$) and can analytically continue into the region $E$ to study Euclidean CFTs. The Jacobian \ref{eq:jacobian} however, as discussed, carries critical information about the causal structure.

Focussing on the Minkowski limit, for  $0<y,\eta<\infty$, one must distinguish between regions $0<\eta<y$ (where \ref{eq:scausal} is non-vanishing) and  $0<y<\eta$ (where \ref{eq:scausal} is vanishing and all points are spacelike separated). Observe that 
asymptotic behavior for $J^{-1}$ differs in these two regions, 
\begin{equation}
	J(y,\eta)^{-1}\simeq \theta(y-\eta) e^{dy/2 +(d-2)\eta/2} + \theta(\eta-y) e^{(d-2)y/2 + d\eta/2}. \label{eq:J}
\end{equation}
The exponents in this equation can be conveniently written in a vector product $\vec\rho\cdot\vec{t}$, where $\vec{t} = (y,\eta)$, and $\vec\rho$ is a $d$ dependant vector. Clearly, $\vec\rho$ is distinct in the two regions of interest:
\begin{equation}
	2\vec{\rho}= 
	\begin{cases}
		(d,d-2),& \text{if } y> \eta\ ;\\
		(d-2,d),& \text{if } y< \eta\ .
	\end{cases}
\end{equation}
This was discussed in \cite{Agarwal:2023xwl} where it was shown that this distinction arises from the way we are required to handle ordering of roots when constructing representations over \textit{semigroups} (see Sec. \ref{sec:groups} for a brief review).

\subsection{Causal Conformal Harmonics}
Let us focus on what follows in the chamber $0<\eta<y<\infty$, corresponding to the $ s $-channel causal scattering region. After resetting the ``ground state energy", i.e., subtracting the constant factor $V_0$, the effective potential $\bar V=V-V_0$ vanishes in the limit $y\pm \eta\rightarrow \infty$. 
This has a positive continuous spectrum, i.e.,   
\begin{equation}
	\left[-\partial_y^2 -\partial_\eta^2 + \bar V(y,\eta)\right] \,{\cal G}_{\widetilde \ell, \widetilde \Delta}(y,\eta)=\bar \nu\,\, {\cal G}_{\widetilde \ell, \widetilde \Delta}(y,\eta)
\end{equation}
with $\bar \nu = \nu -V_0$ positive. The solutions are asymptotically plane-wave, with eigenvalues identified with eigenvalues of  boost $L_{zt}$ and dilatation $D$, 
\begin{equation}
	\bar \nu = k_y^2+k_\eta^2 \equiv  - \widetilde \ell^2  - \widetilde \Delta^2, 
\end{equation}
i.e., $\widetilde \ell$ and $\widetilde \Delta$ purely imaginary.

The solution 
can be formally expressed as 
\begin{equation}
	{\cal G}_{(\widetilde\ell,\widetilde \Delta)}(y,\eta) 
	=  e^{\widetilde \ell \,y - \widetilde \Delta \, \eta  } \sum_{n=0}^\infty \sum_{m=-\infty}^{\infty}A_{n,m} w^{- 2 n }(\sigma/w)^{ -2m }  \label{eq:zsf-d}
\end{equation}
where coefficients $A_{n,m}$ can be found iteratively. The prefactor $e^{\widetilde \ell \,y - \widetilde \Delta \, \eta  }$ follows the convention introduced in \cite{Agarwal:2023xwl}. The sum assures convergence at $\infty$, with $\sigma/w<1$, which is the desired solution for the $ s $-channel scattering region.

It is instructive to compare the eigenvalue for the Casimir, $\nu=\bar \nu+V_0$, to that obtained under  the Euclidean treatment of Eq. \ref{eq:Casimir-value}, i.e., $\nu =C(\ell,\Delta)$. By continuing both $\ell$ and $\Delta$ complex, one finds they can be identified with the corresponding Minkowski values 
\begin{align}
	0<k_y^2 &= - \widetilde \ell^2 =-(\ell+(d-2)/2)^2, \quad {\rm and}\nn 
	0< k_\eta^2 &= - \widetilde \Delta^2 =-(\Delta-d/2)^2
\end{align}

That is, the analytically continued $|\widetilde \ell|$ goes into eigenvalues for the Lorentz boost, $L_{zt}$, and $|\widetilde \Delta|$ into eigenvalues for the dilatation, $D$, in the Minkowski setting, with both $\widetilde \ell$ and $\widetilde \Delta$  purely imaginary. This is indeed our expectation.

\section{Geometry of Radon transform on the disk}\label{app:disc}
This group theoretic formalism for inversion has a natural geometric interpretation. To see this geometry, it is instructive to consider the simplest case of $d = 1$ as it highlights several of the key insights in a way that can be drawn on a page. The relevant group of symmetries is $SO(1,2)$ where the $K$-bi-invariant functions are Legendre $P$ functions whereas the $H$-bi-invariant functions are Legendre $Q$ functions (see App. F of \cite{Agarwal:2023xwl} and Ch. 9 of the series \cite{Vilenkin:1991,Vilenkin1993,Vilenkin:1992}).

\subsection{Euclidean AdS$_2$ ($SO(2,1)/SO(2)$)}
The embedding space $\mathbb{R}^{d,2}$ we are working with has a natural metric that we shall denote by $g_{\alpha\beta} = \text{diag}\{-1,1;-1,1,\cdots,1\}$, with the indices $\alpha,\beta$ in $(-2,-1,0,1,2,\cdots,d-1)$. The indices $(0,\cdots,d-1)$ are the standard Lorentz space indices, whereas $(-2,-1)$ correspond to the embedding space indices required for CFTs (see \cite{Dobrev:1977} for an introduction to the embedding space formalism in the Euclidean setting. More recent references include \cite{Costa:2012cb}.). For the $d = 1$ case in the Minkowski limit, the embedding space is $\mathbb{R}^{1,2}$ and the metric reduces to $(-,+;-)$. 

Let us first considering the Euclidean AdS$_2$ case. We can perform a Wick rotation to go to $\mathbb{R}^{2,1}$ where the metric is $(-,+;+)$. The definition of AdS dictates that, in the embedding space,
\begin{equation}
	-X_{-2}^2 + X_{-1}^2 + X_0^2 = -R^2
\end{equation}
where $R$ is some positive real number we can fix to 1 without loss of generality. The geometry of this manifold is well understood in terms of the Poincar\'e disc model of H$_2$. It is easy to see that the point
\begin{equation}
	\xi_0 = \{1,0;0\}
\end{equation}
is on the manifold and is fixed under the subgroup of rotations $K = SO(2)$ in $(-1,0)$ plane. The group $SO(2,1)$ acts transitively on the disc and maps the point $\xi_0$ into every other point on H$_2$.

Since $G$ admits a Cartan decomposition $G = KAK$, the geometry of Euclidean AdS$_2$ can be identified with the coset manifold $G/K$~\footnote{This is because each point $\xi$ on H$_2$ can be associated with a group element of the form $k(\phi)a(\eta)$ by solving the equation $\xi = ka\cdot\xi_0$. Geometrically, $a$ measures the distance away from the centre of the disc whereas $k$ measures the angle.}. Group elements on the coset can be constructed by making a simple observation: for the subgroup of isometry $K$, it is true that $K^T = K^{-1}$, where the superscript $T$ refers to transpose. One can therefore follow the ``doubling procedure''~\footnote{Strictly speaking, calculations for H$_2$ and AdS$_2$ do not require the doubling procedure and can be studied directly on the manifold. However, it becomes indispensable when studying CFTs in general $d$, since they live on Grassmannians. It is therefore instructive to get used to it in these simpler settings.} outlined in \cite{Agarwal:2023xwl}, and associate with each group element a matrix $I(g)$ such that
\begin{equation}\label{eq:double}
	I(g) = k_1ak_2\cdot k_2^Ta^Tk_1^T = k_1a^2k_1^T
\end{equation}
parametrizes the coset. This gives the equivalent of a polar angle picture where lines of constant $a$ are circles on the disc centred around the origin (see Fig. \ref{fig:disc}(a)).

As discussed in Sec. \ref{sec:review}, the group also admits an Iwasawa decomposition of the form $g = na_Ik$. It leads to a different map on of the coset elements, this time parametrized by $n(c)a_I(\eta_I)$. In the doubling scheme, this leads to the equality
\begin{equation}\label{eq:master}
	I(g) = na_I^2n^T\ = k_1a^2k_1^T\ .
\end{equation}
We refer to Eq. \ref{eq:master} as the master equation. This is the $K$-invariant analogue of Eq. \ref{eq:masterh}. Lines of constant $a_I$ are horocycles of the disk (see Fig. \ref{fig:disc}(b)). A line element on H$_2$ in the two coordinate systems is given by:
\begin{align}
	d\xi^2 &= d\eta^2 + \sinh^2\eta\ d\phi^2\\
	&= d\eta_I^2 + e^{-2\eta_I} dc^2
\end{align}
where $(\phi,\eta)$ parametrize the coset element of the form $ka$ and $(c,\eta_I)$ parametrize $na_I$.

We are now ready to study functions over H$_2$ as outlined in Eq. \ref{eq:s=hm}. Consider a rotationally symmetric function $f$ over H$_2$ such that it depends only on the distance from the centre of the disk. This is parametrized by the one dimensional subgroup $a(\eta)$ in the Cartan decomposition. Such a function should be expressible in terms of a partial wave expansion with a set of basis functions over H$_2$ that are $K$-bi-invariant.

\begin{figure}
	\centering
	\begin{subfigure}{0.3\textwidth}
		\includegraphics[width=\textwidth]{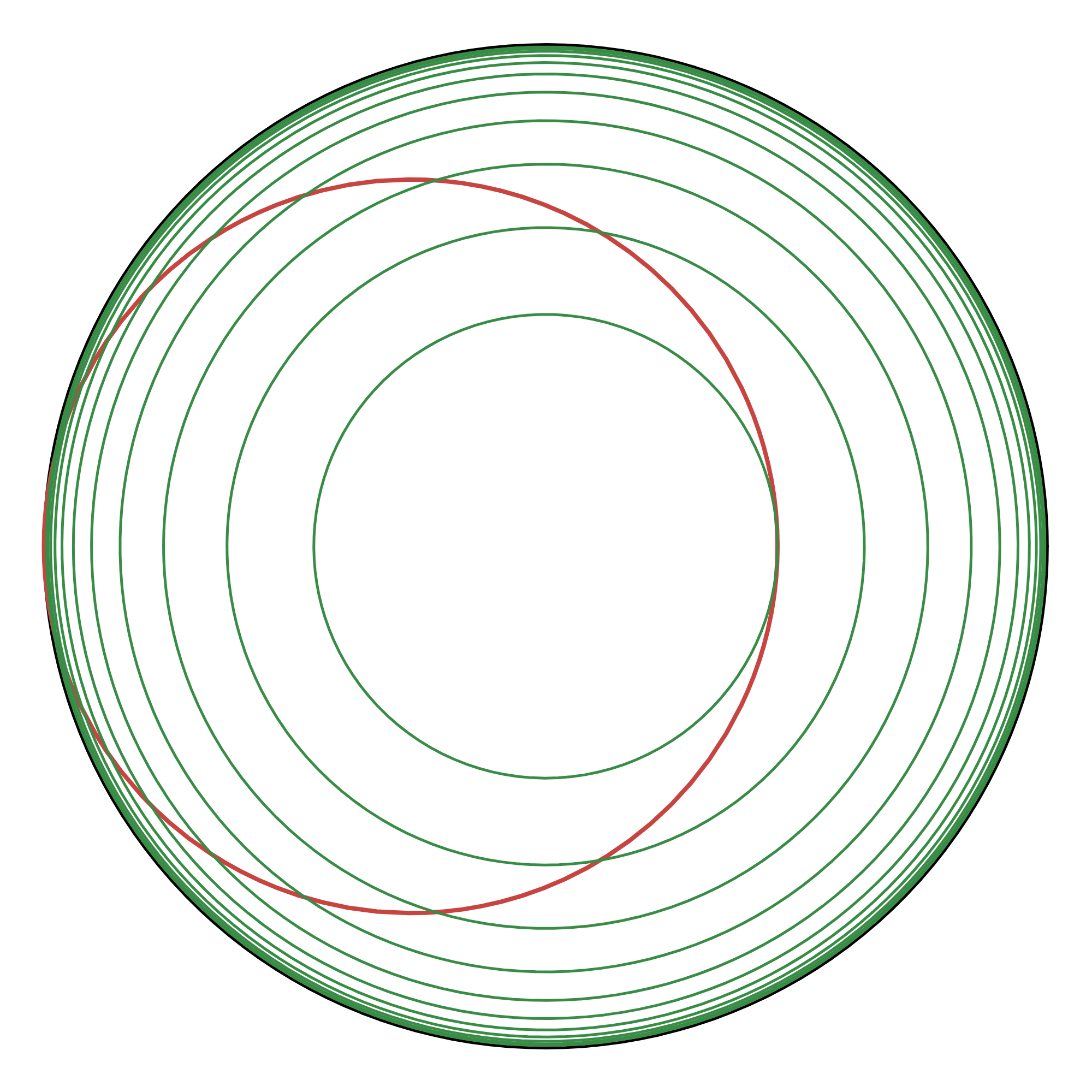}
		\caption{}
	\end{subfigure}
	\caption{The figure shown in red a ``transverse'' slice of the disc H$_2$ over which one integrates to obtain the Abel/Radon transform of the function. The innermost green circle is the line of constant $\eta$ such that $\eta=\eta_I$. When taking the Radon transform, one can integrate over $\eta$ instead of $c$, as long as on maintains $\eta>\eta_I$.}
	\label{fig:radonk}
\end{figure}

The Radon/Abel transform of $f(\eta)$ corresponds to taking a ``transverse'' slice of the function at some constant $\eta_I$ and summing the contribution of all the points on this slice. Equivalently, each slice at constant $\eta_I$ can be parametrized by $\eta$ such that $\eta_I<\eta<\infty$. Geometrically, this is shown in Fig. \ref{fig:radonk}. A change of variable, done by taking a trace of the Master Equation \ref{eq:master}, makes this explicit. We have:
\begin{equation}
	c^2 = 2(\cosh\eta - \cosh\eta_I)e^{-\eta_I}.
\end{equation}
Therefore, the Abel/Radon transform of the function $f(\eta)$ can be written as:
\begin{equation}\label{eq:radonmeasurek}
	F_f(a_I) = \frac{e^{-\eta_I/2}}{\sqrt{2}}\int_{\eta_I}^\infty \frac{f(\eta)}{\sqrt{\cosh\eta-\cosh\eta_I}}\ d(\cosh\eta)\ .
\end{equation}
The partial wave amplitudes are given by a Laplace/Mellin transform of this function:
\begin{equation}
	\hat{f}(\lambda)= \int_{A_I} F_f(a_I)\ e^{\lambda^* a_I}\ da_I\ .
\end{equation}

\bibliographystyle{JHEP}
\bibliography{conformald.bib}

\end{document}